\documentclass[aps,superscriptaddress,preprint,letter,pre] {revtex4}

\usepackage{graphicx}% Include figure files
\usepackage{dcolumn}% Align table columns on decimal point
\usepackage{bm}% bold math
\usepackage{epsf}
\usepackage{amsfonts}
\usepackage{float}
\graphicspath{{figures/}}

\begin{document}

\title{Packing Hyperspheres in High-Dimensional Euclidean Spaces}

\author{Monica Skoge}

\affiliation{\emph{Department of Physics}, \emph{Princeton University},
Princeton, NJ 08544}

\author{Aleksandar Donev}

\affiliation{\emph{Program in Applied and Computational Mathematics}, 
\emph{Princeton University}, Princeton NJ 08544}

\affiliation{\emph{PRISM}, \emph{Princeton University}, Princeton NJ 
08544}

\author{Frank H. Stillinger}

\affiliation{\emph{Department of Chemistry}, \emph{Princeton 
University}, Princeton
NJ 08544}

\author{Salvatore Torquato}

\email{torquato@electron.princeton.edu}

\affiliation{\emph{Program in Applied and Computational Mathematics}, 
\emph{Princeton University}, Princeton NJ 08544}

\affiliation{\emph{PRISM}, \emph{Princeton University}, Princeton NJ 
08544}

\affiliation{\emph{Department of Chemistry}, \emph{Princeton 
University}, Princeton NJ 08544}

\affiliation{\emph{Princeton Center for Theoretical Physics},
\emph{Princeton University}, Princeton NJ 08544}

\newcommand{\changed}[1]{#1}

\begin{abstract}
We present the first study of disordered jammed hard-sphere packings
in four-, five- and six-dimensional Euclidean spaces.  Using a
collision-driven packing generation algorithm, we obtain the first
estimates for the packing fractions of the maximally random jammed
(MRJ) states for space dimensions $d=4$, $5$ and $6$ to be $\phi_{MRJ}
\simeq 0.46$, $0.31$ and $0.20$, respectively.  To a good
approximation, the MRJ density obeys the scaling
form $\phi_{MRJ}= c_1/2^d+(c_2 d)/2^d$,
where $c_1=-2.72$ and $c_2=2.56$, which appears to be consistent
with high-dimensional asymptotic limit, albeit with different coefficients.
Calculations of the pair correlation function $g_{2}(r)$ and structure factor $S(k)$ for
these states show that short-range ordering appreciably decreases with
increasing dimension, consistent with a recently proposed
``decorrelation principle,'' which, among othe things,
states that unconstrained correlations
diminish as the dimension increases and vanish entirely in the limit
$d \rightarrow \infty$. As in three dimensions (where $\phi_{MRJ}
\simeq 0.64$), the packings show no signs of crystallization, are
isostatic, and have a power-law divergence in $g_{2}(r)$ at contact
with power-law exponent $\simeq 0.4$.  Across dimensions, the
cumulative number of neighbors equals the kissing number of the
conjectured densest packing close to where $g_{2}(r)$ has its first
minimum.  Additionally, we obtain estimates for the freezing and
melting packing fractions for the equilibrium hard-sphere fluid-solid
transition, $\phi_F \simeq 0.32$ and $\phi_M \simeq 0.39$,
respectively, for $d=4$, and $\phi_F \simeq 0.19$ and $\phi_M \simeq
0.24$, respectively, for $d=5$. Although our results indicate the
stable phase at high density is a crystalline solid, nucleation
appears to be strongly suppressed with increasing dimension.
\end{abstract}

\maketitle

\section{Introduction}

Hard-sphere systems are model systems for understanding the
equilibrium and dynamical properties of a variety of materials,
including simple fluids, colloids, glasses, and granular media.  The
hard-sphere potential is purely repulsive; it is infinite when two
spheres overlap, but otherwise zero.  Despite the simplicity of the
potential, hard-sphere systems exhibit rich behavior: they undergo a
fluid-solid phase transition and can exhibit glassy behavior.  Of
particular recent interest are (nonequilibrium) disordered jammed
packings of hard spheres and their statistical and mechanical
properties, such as the maximally random jammed (MRJ) state~\cite{rcp,
kansal}, pair correlations~\cite{aleks}, isostaticity~\cite{aleks},
and density fluctuations~\cite{alekshyper}.  Such packings have been
intensely studied computationally in two and three
dimensions~\cite{finney, bennett, tobochnik, zinchenko, rcp, kansal,
rintoul, aleks, alekshyper, ohern, binarydisk} and in this paper we
extend these studies to four, five and six dimensions.

A hard-sphere packing in $d$-dimensional Euclidean space
$\mathbb{R}^d$ is an arrangement of congruent spheres, no two of which
overlap.  As in a variety of interacting many-body
systems~\cite{chaikin}, we expect studies of hard-sphere packings in
high dimensions to yield great insight into the corresponding
phenomena in lower dimensions.  Analytical investigations of
hard-spheres can be readily extended into arbitrary spatial
dimension~\cite{virial234, virial234too, EOSlubanmichels, virial56, Virial4_EvenDimensions,
frisch, parisi, finken, philipse, torquatobook, hyperuniformity, Pa06,         
decorrelation, torquato15, To06} and high dimensions can therefore be used as a
stringent testing ground for such theories.   Along these lines and of
particular interest to this paper, predictions have been made about
long-wavelength density fluctuations~\cite{hyperuniformity} and
decorrelation~\cite{decorrelation,torquato15} in disordered hard-sphere packings
in high dimensions.  Additionally, the optimal packing of hard spheres
in high dimensions is also of interest in error-correcting codes in
communications theory~\cite{conway}.

Our focus in this paper will be the study of hard-sphere packings in
four, five and six dimensions.  Specifically, we consider both
equilibrium packings for $d=4$ and $d=5$ and nonequilibrium
packings representative of the maximally random jammed state for
$d=4$, $d=5$ and $d=6$.

Equilibrium thermodynamic properties of hard-sphere packings for $d=4$
and $d=5$ have been studied both computationally  and with approximate
theories \cite{EOSlubanmichels,finken,Lue}.  For the low-density fluid, lower-order virial coefficients,
$B_2$, $B_3$, and $B_4$, are known exactly for arbitrary
dimensionality~\cite{virial234, virial234too,Virial4_EvenDimensions}.  Higher-order virial
coefficients have been calculated by Monte Carlo simulation, $B_5$,
$B_6$ and $B_7$ for both $d=4$ and $d=5$~\cite{virial56} and
$B_8$ for $d=4$~\cite{virial56}, and analytically \cite{Lyberg,Virial89_HighDimensions}.  
The pair correlation function for equilibrium fluids has been studied and a decrease 
in ordering with increasing dimension was readily apparent~\cite{bishop}.  Hard-sphere
systems have been shown to undergo a (first-order) fluid-solid phase
transition by numerical simulations for $3 \leq d \leq
5$~\cite{michelstrapp} and with approximate theories for $d$ as high
as $50$~\cite{finken}.  The freezing points for $d=4$ and $d=5$ were
estimated numerically to occur at packing fractions 
$\phi_F \approx 0.5 \phi_{max} = 0.31$ and $\phi_F
\approx 0.4 \phi_{max} = 0.19$, respectively, and it was conjectured that
freezing occurs at lower packing fractions relative to close packing
as the dimension increases~\cite{michelstrapp}.  The {\it packing fraction}
$\phi$ is the fraction of
space $\mathbb{R}^d$ covered by the spheres, i.e.,
\begin{equation}
\phi=\rho v_1(R),
\label{phi}
\end{equation}
where $\rho$ is the number density,
\begin{equation}
v_1(R) = \frac{\pi^{d/2}}{\Gamma(1+d/2)} R^d
\label{v(R)}
\end{equation}
is the volume of a $d$-dimensional sphere of radius $R$, and
$\Gamma(x)$ is the gamma function \cite{torquatobook}.

At sufficiently large densities, the packing of spheres with the
highest jamming density has the greatest entropy because the
free-volume entropy dominates over the degeneracy entropy.  
Therefore, the high-density equilibrium phase corresponds to the
optimal packing, {\em i.e.,} maximal density.  The densest packing
for $d=3$ was recently proven by Hales~\cite{hales} to be attained by
the FCC lattice with packing fraction $\phi_{max} = \pi/\sqrt{18} =
0.7404 \dots$.  The kissing number $Z$, the number of spheres in
contact with any given sphere, for the FCC lattice corresponds to the
maximal kissing number $Z_{max} = 12$ for $d=3$.  One of the
generalizations of the FCC lattice to higher dimensions is the $D_d$
checkerboard lattice, defined by taking a cubic lattice and placing
spheres on every site at which the sum of the lattice indices is even
({\em i.e.,} every other site).  The densest packing for $d=4$ is
conjectured to be the $D_4$ lattice, with packing fraction $\phi_{max}
= \pi^2/16 = 0.6168 \ldots$ and kissing number $Z = Z_{max} =
24$~\cite{conway}, which is also the maximal kissing number in
$d=4$~\cite{musin}.  For $d=5$, the densest packing is conjectured to
be the $D_5$ lattice, with packing fraction $\phi_{max} =
2\pi^2/(30\sqrt{2}) = 0.4652 \ldots$ and kissing number $Z =
40$~\cite{conway}.  For $d=6$, the densest packing is conjectured to
be the ``root'' lattice $E_6$, with density $\phi_{max} =
3\pi^3/(144\sqrt{3}) = 0.3729 \ldots$ and kissing number $Z =
72$~\cite{conway}.  The maximal kissing numbers $Z_{max}$ for $d=5$
and $d=6$ are not known, but have the following bounds: $40 \leq
Z_{max} \leq 46$ for $d=5$ and $72 \leq Z_{max} \leq 82$ for
$d=6$~\cite{conway}.  In very high dimensions, it has been suggested
that random packings of spheres might have a higher density than
ordered packings, enabling the intriguing possibility of disordered
ground states and hence thermodynamic glass
transitions~\cite{decorrelation}; see also Ref.~\cite{torquato15}.

\begin{figure}
\begin{center}
\includegraphics*[width=0.7\columnwidth,keepaspectratio]{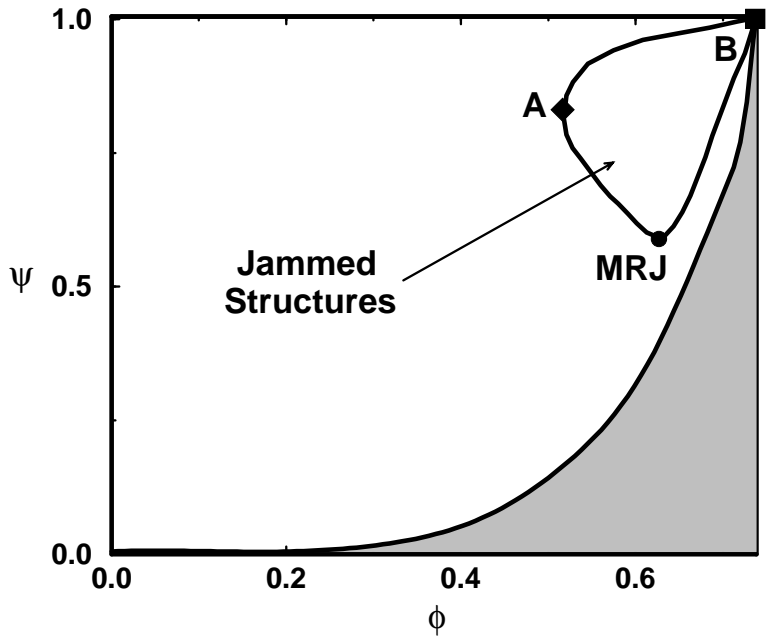}
\caption{A highly schematic plot of the subspace in the
density-disorder $\phi-\psi$ plane, where strictly jammed
three-dimensional packings exist, as adapted from Ref.~\cite{rcp}.
Point A corresponds to the lowest-density jammed packing, and it is
intuitive to expect that a certain ordering will be needed to produce
low-density jammed packings.  Point B corresponds to the most dense
jammed packing, which is also expected to be the most ordered.  Point
MRJ represents the maximally random jammed state.  The jamming region
in the $\phi$-$\psi$ plane will of course depend on the jamming
category.  The gray region is devoid of hard-sphere
configurations.  }
\label{phipsi}
\end{center}
\end{figure}

Equilibrium hard-sphere systems for $d=2$ and $d=3$ crystallize into
ordered packings upon densification. However, for $d=3$, it has been
found both experimentally~\cite{Sc69} and
computationally~\cite{rintoul, rcp, aleks} that if the system is
densified sufficiently rapidly, the system can be kept
out-of-equilibrium and can jam in a disordered state. A jammed packing
is one in which the particle positions are fixed by the
impenetrability constraints and boundary conditions, despite thermal
or mechanical agitation of the particles or imposed boundary
deformations or loads. Depending on the boundary conditions, different
jamming categories can be precisely defined, including local,
collective and strict jamming \cite{linprogramming, category1,
category2}.  The density of disordered collectively jammed hard-sphere
packings for $d=3$ is around $\phi \simeq 0.64$ for a variety of
packing-generation protocols and has traditionally been called random
close packing (RCP) \cite{torquatobook}.  However, Ref.~\cite{rcp} showed
that RCP is ill-defined because ``random'' and ``close packed'' are at
odds with one another and the precise proportion of each of these
competing effects is arbitrary. Therefore, Ref.~\cite{rcp} introduced
the concept of the maximally random jammed (MRJ) state to be the most
disordered jammed packing in the given jamming category.  This
definition presupposes an order metric $\psi$ can be defined such that
$\psi = 1$ corresponds to the most ordered ({\em i.e.,} crystal)
packing and $\psi = 0$ corresponds to the most disordered packing, in
which there are no spatial correlations.  Figure~\ref{phipsi} from
Ref.~\cite{rcp} shows where MRJ lies on a schematic diagram of the
space of jammed packings in the density-disorder $\phi$-$\psi$ plane.

In this paper, we numerically study MRJ packings of hard spheres for
$d=4$, $5$ and $6$ that are at least collectively jammed and report
the first estimates of the packing fractions of the MRJ
states~\cite{rcp} in these dimensions to be $\phi_{MRJ} \simeq 0.46$,
$0.31$ and $0.20$, respectively.  We find that short-range ordering
exhibited by $g_{2}(r)$ and $S(k)$ appreciably diminishes with
increasing dimension, consistent with a recently proposed
``decorrelation principle'' stating that unconstrained spatial
correlations vanish asymptotically in high dimensions and
that the $n$-particle correlation function
$g_n$ for any $n \ge 3$ can be inferred entirely from a knowledge
of the number density $\rho$ and the pair correlation function $g_2({\bf r})$ 
~\cite{decorrelation,footnote}.  We
also explore equilibrium properties, in particular the fluid-solid
phase transition, for $d=4$ and $d=5$, and find a decreased tendency
to crystallize with increasing dimension.

This paper is organized as follows: Section II explains the simulation
procedure, Section III gives equilibrium results for $d=4$ and $d=5$,
Section IV gives results for disordered jammed packings for $d=4$, $5$
and $6$, and Section V summarizes and discusses our results.

\section{Simulation Procedure}

We use event-driven molecular dynamics and a modified
Lubachevsky-Stillinger (LS) algorithm~\cite{LS1}, as in
Ref.~\cite{HSalg}, to produce collectively-jammed hard-sphere
packings.  As in Ref.~\cite{HSalg}, our algorithm uses
periodic boundary conditions applied to a hypercubic cell, in which a
fundamental cell containing $N$ spheres is periodically replicated to
fill all of Euclidean space.  We also use the cell method, in which
the computational domain is divided into cubic cells and only
neighboring cells are checked when predicting collisions for a given
sphere.  Since the number of neighboring cells, as well as the number
of spheres per cell, increases considerably with increasing dimension,
working in high dimensions is computationally intensive.
Additionally, eliminating excessive boundary effects requires on the
order of ten sphere diameters per simulation box length, {\em i.e.,}
on the order of $N=10^d$ spheres.  Due to the increasing computational
load with increasing dimension, we cannot yet study $d > 6$.
Implementing the near-neighbor list (NNL) techniques from
Ref.~\cite{HSalg}, as well as parallelization, are necessary in order
to study higher dimensions. Dimension-independent C++ codes used to
generate the data in this paper can be downloaded at {\tt
http://cherrypit.princeton.edu/Packing/C++/}.

Starting from a Poisson distribution of points, the points grow into
nonoverlapping spheres of diameter $D$ at an expansion rate $\gamma =
dD/dt$, while the positions of the spheres evolve in time according to
Newtonian mechanics, augmented with energy non-conserving collisions.
Spheres receive an extra energy boost after the collision due to the positive 
expansion rate. In practice, the starting configurations for our packing 
algorithm are low density random-sequential-addition packings of
spheres~\cite{torquatobook}.  As the density increases, statistics,
such as pressure, are collected.  In the limit $\gamma \rightarrow 0$,
the system is in equilibrium; for small but nonzero $\gamma$, the
system is in quasi-equilibrium; and for large $\gamma$, the system is
out of equilibrium.  Eventually, a jammed state with diverging
collision rate is reached.  For studies of amorphous jammed packings,
the expansion must be initially fast to suppress crystallization and
maximize disorder, but at sufficiently high pressure, the expansion
rate must be slow enough to allow local particle rearrangements
necessary to achieve jamming~\cite{aleks}.

\section{Equilibrium and Metastable Properties}

The temperature in equilibrium systems of hard spheres is a trivial
variable; {\em i.e.,} it does not affect equilibrium configurational
correlations, leaving only one independent thermodynamic state
variable, which can be taken to be either the reduced pressure $p =
PV/Nk_BT$ or the density $\phi$, related through the equation of state
(EOS).  Hard-sphere systems undergo a (first-order) fluid-solid phase
transition, characterized by a melting point, {\em i.e.,} the density
at which the crystal thermodynamically begins to melt, and a freezing
point, {\em i.e.,} the density at which the fluid thermodynamically
begins to freeze.  Equilibrium properties, such as the melting and
freezing points, are studied here using small expansion rates ($\gamma
= 10^{-5}-10^{-9}$) and periodic rescaling of the average sphere
velocity to one, such that the total change in kinetic energy of the
system, due to the collisions between growing spheres, was kept small.
Strictly speaking, a positive growth rate yields nonequilibrium packings
but equilibrium packings result as the growth rate
tends to zero. The packings were ``equilibrated" by verifying that 
orders of magnitude of change in the expansion rate did not change the 
resulting equation of state. 
In this section we only consider four and five dimensions due to
(presently) prohibitive computational costs for higher dimensions.

\begin{figure}[bthp]
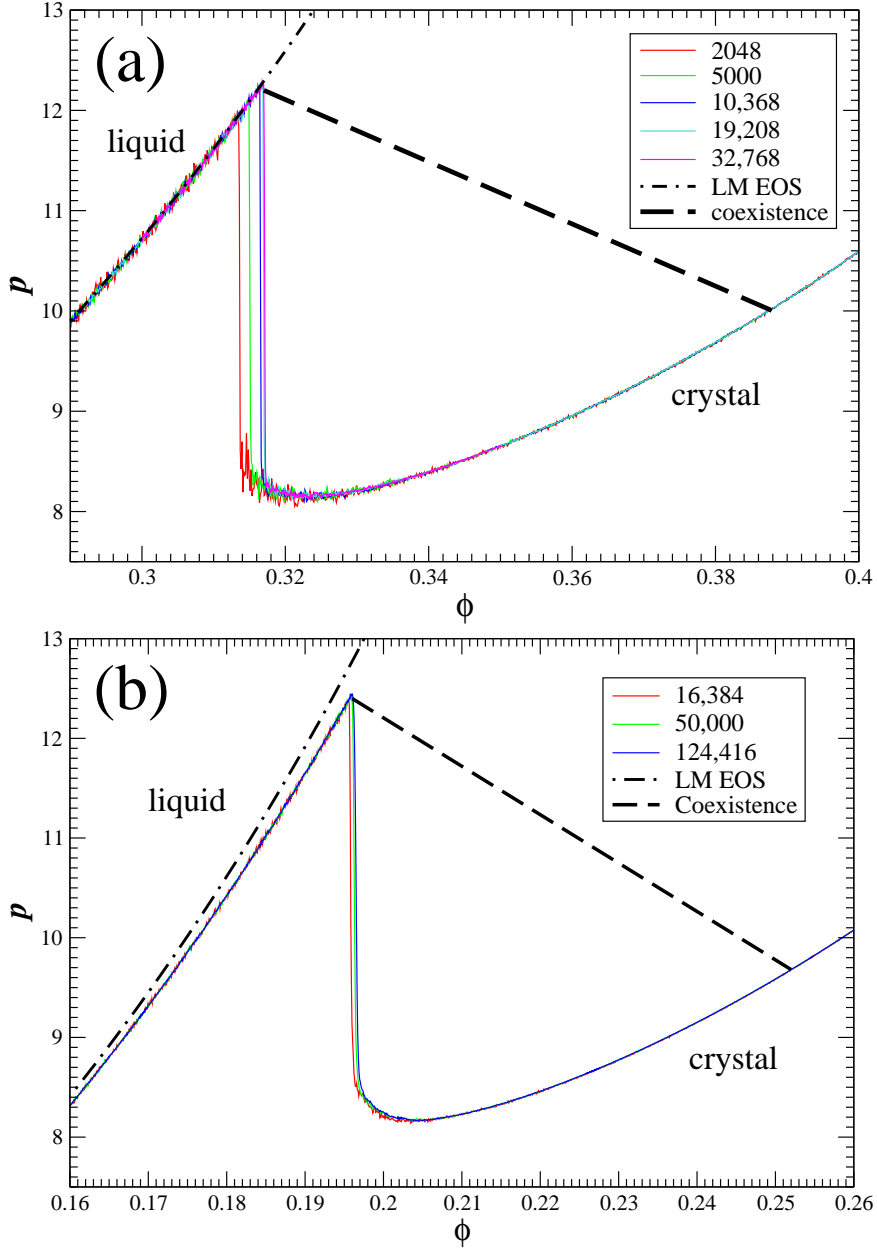

\begin{center}
\includegraphics*[width=0.7\columnwidth,keepaspectratio]{melt6.eps}
\includegraphics*[width=0.7\columnwidth,keepaspectratio]{melt5D.eps}
\caption{(Color online) Reduced pressure $p$ as a function of density
$\phi$, for a range of system sizes (see legend), for (a) $d=4$
systems of spheres, initially in a $D_4$ lattice, and negative
expansion rate $\gamma = -10^{-6}$ and (b) $d=5$ systems of spheres,
initially in a $D_5$ lattice, and negative expansion rate $\gamma =
-10^{-5}$.  $N$ was chosen to make a perfect $D_d$ lattice with
periodic boundary conditions, {\em i.e.,} $N = (2n)^d/2$ for $n~
\epsilon ~\mathbb{Z}$.  Also plotted is the theoretical prediction of
Luban and Michels (LM) for the equation of
state~\cite{EOSlubanmichels}.  Curves for larger system sizes lie
farther to the right.  }
\label{melt}
\end{center}
\end{figure}

Figure~\ref{melt} shows the reduced pressure $p$ as a function of
density $\phi$ for (a) simulations of $d=4$ systems of spheres placed
in a $D_4$ lattice with negative expansion rate $\gamma = -10^{-6}$
and (b) simulations of $d=5$ systems of spheres placed in a $D_5$
lattice with negative expansion rate $\gamma = -10^{-5}$.  The
pressure initially follows the (lower) crystal branch, until the
system becomes mechanically unstable and jumps onto the (higher) fluid
branch.  Also plotted is the theoretical prediction of Luban and
Michels (LM) for the equation of state~\cite{EOSlubanmichels}, which
agrees well with our numerical results for the fluid branch for $d=4$,
but less so for $d=5$.  It is a computational observation that
crystals become mechanically unstable, giving rise to a sudden jump in
pressure, at a density close to the freezing point~\cite{melting1,
melting2}.  Such ``superheating'' (undercompression) is most
likely due to the difficulty of achieving coexistence in finite
systems, although we are not aware of a theoretical analysis.  From
the results in Fig.~\ref{melt}, we estimate the freezing points for
$d=4$ and $d=5$ to be $\phi_F \simeq 0.31-0.32$ and $\phi_F \simeq
0.19-0.20$, respectively.

The melting points for $d=4$ and $d=5$ can also be estimated from the
data in Fig.~\ref{melt}.  Since throughout the coexistence region the
fluid and solid have the same absolute pressure $P$, the melting
density can be estimated as the density on the crystal branch with the
same absolute pressure $P$ as that at the freezing point.  The
coexistence region is plotted in Fig.~\ref{melt} and the melting
packing fractions for $d=4$ and $d=5$ are estimated to be $\phi_M
\simeq 0.38-0.40$ and $\phi_M \simeq 0.24-0.26$, respectively.  We
also observe that the reduced pressure at the freezing point is $p_F
\simeq 12$ in both $d=4$ and $d=5$, which agrees with the reduced
pressure at the freezing point for $d=3$, $p_F \simeq 12.3$, obtained
from free energy calculations~\cite{frenkel}.

\begin{figure}[H]
\begin{center}
\includegraphics*[width=0.95\columnwidth,keepaspectratio]{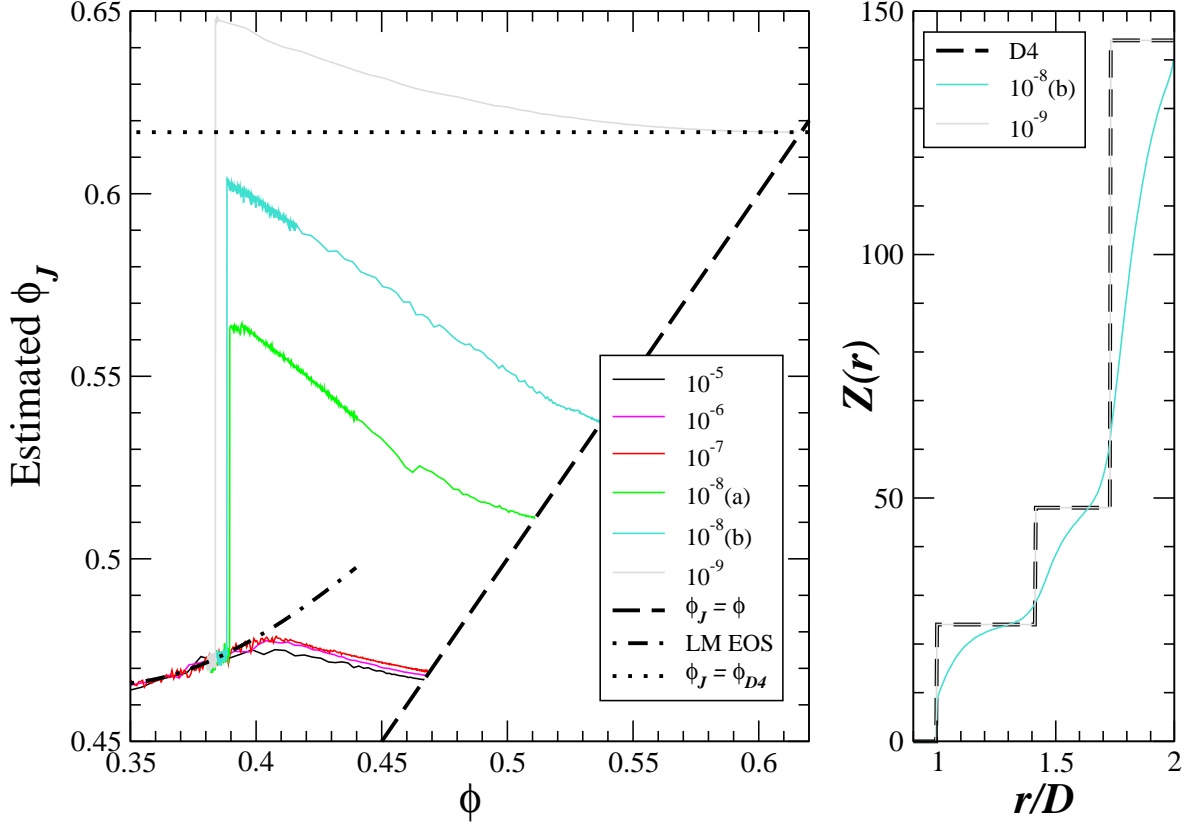}
\caption{(Color online) Left panel: The estimated jamming packing
fraction $\tilde \phi_J$ as a function of density $\phi$ for systems
of $648$ spheres for $d=4$ with various expansion rates (see legend
and note that there are two samples labeled (a) and (b) for
$\gamma=10^{-8}$). For the curves showing no partial crystallization
({\em i.e.}, $\gamma = 10^{-5}$, $10^{-6}$, and $10^{-7}$), curves
with smaller expansion rates have larger peak heights.  For the curves
that show partial crystallization ({\em i.e.} $\gamma = 10^{-8}$ (a
and b) and $10^{-9}$), curves with smaller expansion rate lie farther
to the left.  Right panel: The cumulative coordination $Z(r)$ ({\em
i.e.,} the number of contacts) for the perfect $D_4$ lattice and for
the partially crystallized packings at $p>10^{12}$ obtained for
expansion rates $\gamma = 10^{-8}$ and $\gamma = 10^{-9}$.  The
jamming packing fraction for the $\gamma = 10^{-8}$ packing is $\phi =
0.511$, and the jamming packing fraction for the $\gamma = 10^{-9}$
packing agreed up to 12 significant figures with the density of the
$D_4$ lattice, $\phi = \pi^2/16 \simeq 0.617$.  }
\label{freeze}
\end{center}
\end{figure}

The melting point was also estimated for $d=4$ (higher dimensions are
presently too computationally demanding) by slowly densifying a system
of spheres, initially a fluid, and looking for the onset of partial
crystallization, again by monitoring the reduced pressure $p$ as a
function of density $\phi$.  Due to the difficulty of observing
coexistence in finite systems and the relatively high activation
barrier, simulated hard-sphere systems become
``supercooled'' (overcompressed) and nucleation does not occur until
the melting density is surpassed.  Consequently, the density at which
partial crystallization appears for sufficiently slow expansion
provides a reasonable estimate for the melting density.  Near jamming
the reduced pressure is asymptotically given by the free-volume
equation of state~\cite{phiJ},
\begin{equation}
p = \frac{PV}{Nk_BT} = \frac{1}{\delta} = \frac{d}{1-\phi/\phi_J},
\end{equation}
which can be inverted to give an estimate $\tilde \phi_J$ of the
jamming density,
\begin{equation}
\tilde \phi_J = \frac{\phi}{1-d/p}.
\end{equation}
Since the pressure increases very rapidly near jamming, it is more
convenient to plot the estimated jamming density $\tilde \phi_J
(\phi)$ instead of the pressure $p(\phi)$, as shown in
Fig.~\ref{freeze} for a system of $648$ spheres in $d=4$.  In such a
plot, the onset of partial crystallization causes a dramatic jump in
$\tilde \phi_J (\phi)$, as the jamming density of the crystal is much
higher than the jamming density of a disordered packing.  The
intersection of the curves with the line $\tilde \phi_J (\phi) = \phi$
gives the final jamming density. Sufficiently fast expansion
suppresses crystallization and leads to packing fractions around
$0.45-0.47$.  Slower expansion allows for partial crystallization,
typically around $\phi_M \simeq 0.38-0.39$, which is our rough
estimate of the melting point, in agreement with our estimate from the
results in Fig.~\ref{melt}.  More accurate estimates can only be
obtained using free-energy calculations.  Since crystallization is a
nucleated process, it is not surprising that the same expansion
rates $\gamma$ can crystallize at different packing fractions and onto
different crystal branches, {\em e.g.}  $\gamma = 10^{-8}$ (a) and (b)
in Fig.~\ref{freeze}.

To determine whether the crystallized packings were forming a $D_4$
lattice, the conjectured densest packing in four dimensions, we
computed the average cumulative coordination number $Z(r)$, which is
the average number of sphere centers within a distance $r$ from a
given sphere center.  The inset to Fig.~\ref{freeze} shows $Z(r)$ for
a perfect $D_4$ lattice and for the crystallized packings with $\gamma
= 10^{-8}$ and $\gamma = 10^{-9}$ (corresponding colors represent the
same packing).  The sharp plateaus for the $D_4$ lattice correspond to
the coordination shells and the number of spheres in the first shell
is the kissing number $Z_{max} = 24$.  The packing shown with $\gamma =
10^{-9}$ formed a perfect $D_4$ lattice.  The packing shown with $\gamma =
10^{-8}$ partially crystallized with a final density of $\phi \simeq
0.511$.

\begin{figure}
\begin{center}
\includegraphics*[width=0.95\columnwidth,keepaspectratio]{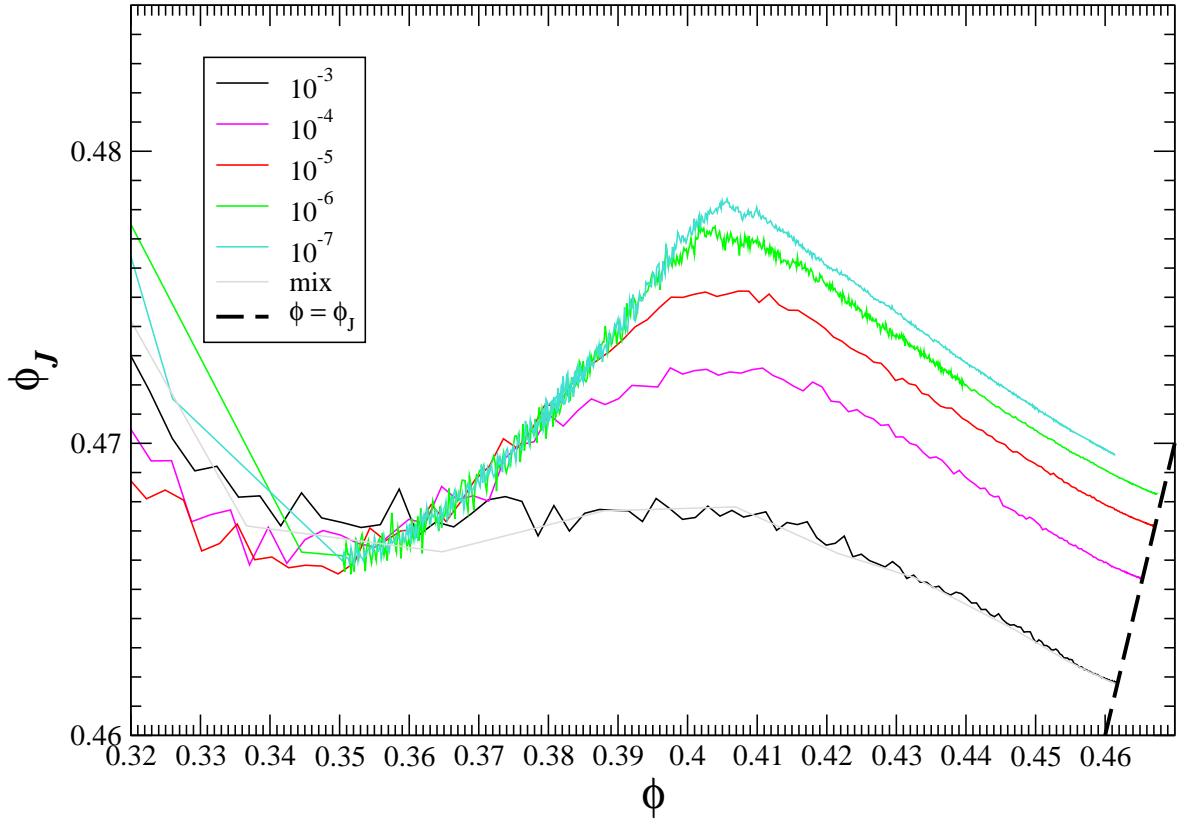}
\caption{The estimated jamming packing fraction $\tilde \phi_J$ as a
function of density $\phi$ for a system of $10,000$ spheres for $d=4$
with various expansion rates.  Curves with smaller expansion rates
have larger peak heights.  The curve labeled ``mix'' corresponds to
the following sequence of expansion rates: $\gamma = 10^{-2}$ until
$p=10$, $\gamma = 10^{-3}$ until $p=10^4$, $\gamma = 10^{-4}$ until
$p=10^6$, and $\gamma = 10^{-5}$ until $p=10^{12}$.  }
\label{10kfreeze}
\end{center}
\end{figure}

Figure~\ref{10kfreeze} shows the estimated jamming packing fraction
$\tilde \phi_J$, as in Fig.~\ref{freeze}, but for a system of $10,000$
spheres, instead of $648$ spheres, in four dimensions.  In contrast
to the $648$ sphere system, there is no sign of partial
crystallization for the $10,000$-sphere system.  In fact, molecular
dynamics was performed at packing fractions of $\phi \simeq 0.38-0.42$
for 10 million collisions per sphere and there was no significant drop
in pressure indicative of partial crystallization.  The curves in
Figs.~\ref{freeze} and~\ref{10kfreeze} exhibit a bump around $\phi_G
\simeq 0.41$, suggesting a kinetic transition from the fluid branch to
a glassy branch.

\begin{figure}[bthp]
\begin{center}
\includegraphics*[width=0.95\columnwidth,keepaspectratio]{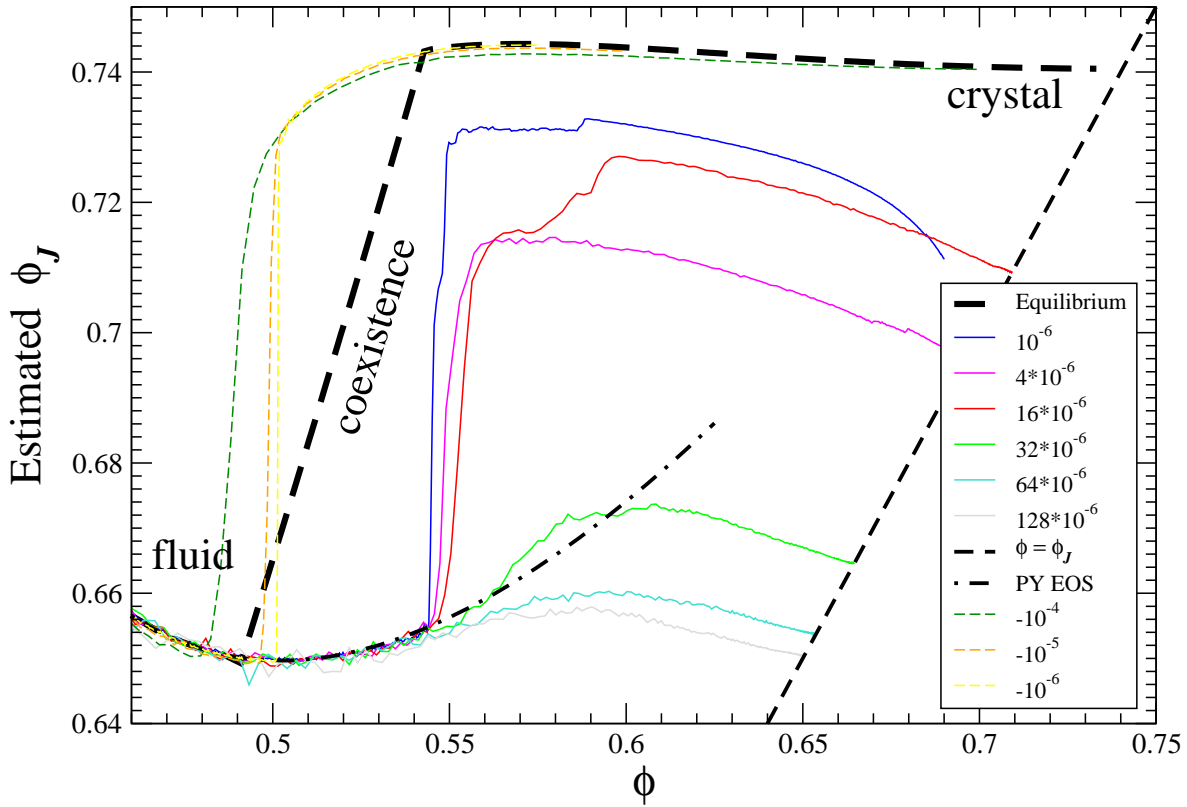}
\caption{(Color online) The estimated jamming packing fraction $\tilde
\phi_J$ as a function of packing fraction $\phi$ for $d=3$.  Shown are
systems of $4096$ spheres with various expansion rates and systems of
$10,976$ spheres placed in an FCC lattice with negative expansion
rates $\gamma = -10^{-4}$, $-10^{-5}$, and $-10^{-6}$ (last three
curves). Also plotted are approximations to the equilibrium EOS for
the fluid phase, the coexistence region, and the crystal
phase~\cite{speedy}, as well as the Percus-Yevick (PY) EOS for the
fluid phase.  Compare this figure to the curves shown in
Figs.~\ref{freeze} and~\ref{10kfreeze}.  For the curves showing no
partial crystallization ({\em i.e.}, $\gamma = 32\times10^{-6}$,
$64\times10^{-6}$, and $128\times10^{-6}$), curves with smaller
expansion rates have larger peak heights.  For the curves that show
partial crystallization ({\em i.e.}, $\gamma = 10^{-6}$,
$4\times10^{-6}$, and $16\times10^{-6}$), curves with smaller expanion
rates lie farther to the left.  For the melting curves ({\em i.e.},
$\gamma = -10^{-4}$, $-10^{-5}$, and $-10^{-6}$), curves with smaller
compression rates lie farther to the right.  }
\label{3D}
\end{center}
\end{figure}

Figure~\ref{3D} shows the estimated jamming packing fraction $\tilde
\phi_J$ for systems of spheres for $d=3$ with various positive and
negative expansion rates, for comparison with the results for $d=4$
and $d=5$ in Figs.~\ref{melt},~\ref{freeze} and~\ref{10kfreeze}.  The
locations of the freezing and melting points in $d=3$ have been
determined from free-energy calculations~\cite{frenkel} and good
approximations to the EOS for both the fluid and crystal
phases are known~\cite{speedy}.  Our estimates of the freezing and
melting points as the densities at the onset of melting of a diluted
crystal or of partial crystallization of a densified fluid,
respectively, compare favorably to the true values computed from
free-energy calculations in $d=3$.  The bump around $\phi_G \simeq
0.59$, analogous to the bump in Fig.~\ref{10kfreeze} around $\phi_G
\simeq 0.41$, is often cited as the approximate location of the
``kinetic'' glass transition~\cite{chaikin2}.  Comparing
Figs.~\ref{10kfreeze} and~\ref{3D} reveals that the melting point and
suggested kinetic glass transition are closer for $d=4$ than for
$d=3$, which is a possible reason why there is a lower tendency to
crystallize for $d=4$ than for $d=3$.  Similar results have been
observed for binary hard disks, a model glass
former~\cite{binarydisk}.

\section{Disordered Jammed Packings}

Packings representative of the maximally random jammed (MRJ)
state are produced by a combination of expansion rates.  The expansion
rate must be initially high (compared to the average thermal velocity)
to suppress crystallization and produce disordered configurations that
are trapped in the neighborhood of a jammed packing.  Near the jamming
point, the expansion rate must be sufficiently slow to allow for
particle readjustments necessary for collective jamming.
Figure~\ref{10kfreeze} shows the final jamming packing fractions of
packings created using a variety of expansion rates, as the packing
fraction at which the curves intersect the line $\tilde \phi_J=\phi$.
We see that by increasing the expansion rate, we attain packings with
lower jamming packing fractions.  

By comparing Fig.~\ref{10kfreeze} and
to the analogous plot for a $d=3$ system (Fig.~\ref{3D}),
where it is widely accepted that $\phi_{MRJ} \simeq
0.64-0.65$~\cite{rcp, kansal}, we estimate the MRJ density for $d=4$
to be $\phi_{MRJ} \simeq 0.460 \pm 0.005$.  A more accurate
calculation of $\phi_{MRJ}$ demands a better theoretical understanding
of order metrics and how the expansion rate in the algorithm affects
the ordering in the produced packings; statistical errors are smaller
than the effect of the packing-generation protocol.  Systematic
investigation of different protocol parameters, as done for $d=4$ in
Fig.~\ref{10kfreeze}, is currently too computationally intensive in
higher dimensions.  Reasonable estimates of $\phi_{MRJ}$ for 
both $d=5$ and $d=6$  are obtained  using the following
less computationally intensive procedure. First, the system
of spheres is expanded, starting from zero
initial kinetic energy ($T=0$), until it reached a high pressure (say,
$p=100-1000$). Then the system is slowly expanded ($\gamma = 10^{-5}-10^{-3}$)
and periodically cooled to $k_BT = 1$ until a
very high pressure (say, $p=10^{12}$) is attained.  The resulting packings are
approximately collectively jammed, as demonstrated by very large
relaxation times for the pressure during long molecular dynamics
runs~\cite{aleks}.  Using this method we estimate the MRJ density for
$d=5$ to be $\phi_{MRJ} \simeq 0.310 \pm 0.005$ and for $d=6$ to be
$\phi_{MRJ} \simeq 0.200 \pm 0.01$. 

 The MRJ packing fractions
as well as important equilibrium packing fractions are summarized
in Table \ref{table}. It is useful to compare the MRJ packings fractions
for $3 \le d \le 6$ to recent estimates of the {\it saturation} packing fraction
$\phi_s$ for the random sequential addition (RSA) packing of hard spheres
obtained by Torquato, Uche and Stillinger \cite{To06} in
corresponding dimensions, which were shown to be nearly hyperuniform \cite{hyperuniformity}.
These authors found that
$\phi_s=0.38278   \pm 0.000046, 0.25454   \pm 0.000091, 0.16102   \pm 0.000036$
and $0.09394   \pm 0.000048$ for $d=3,4,5$ and $6$, respectively.
The nonequilibrium RSA packing is  produced by
randomly, irreversibly, and sequentially placing nonoverlapping spheres into
a volume. As the process
continues, it becomes more difficult to find available
regions into which the spheres can be added. Eventually,
in the saturation (infinite-time) limit, no further additions are possible,
and the maximal achievable packing fraction is the saturation value
$\phi_s$ [see Ref. \cite{torquatobook} and references therein]. 
As expected, the 
RSA saturation packing fraction in dimension $d$ is substantially smaller than
the corresponding MRJ value because, unlike the latter packing,
the particles cannot rearrange.

\begin{table*}
\begin{center}
\begin{tabular}{|c|c|c|c|c|}
\hline
Packing &  &  &  &  \\
fraction & $d=3$ & $d=4$ & $d=5$ & $d=6$ \\
\hline\hline
$\phi_F$ & $0.494$~\cite{torquatobook, frenkel} & $0.32 \pm 0.01^*$ & $0.19 \pm 0.01^*$ & - \\
$\phi_M$ & $0.545$~\cite{torquatobook, frenkel} & $0.39 \pm 0.01^*$ & $0.24 \pm 0.01^*$ & - \\
$\phi_{MRJ}$ & $0.645 \pm 0.005$~\cite{kansal} & $0.46 \pm 0.005^*$ & $0.31 \pm 0.005^*$ & $0.20 \pm 0.01^*$ \\
$\phi_{max}$ & $0.7405 \ldots$~\cite{hales} & $0.6169 \ldots$~\cite{conway} & $0.4652 \dots$~\cite{conway} & $0.3729 \dots$~\cite{conway}\\
\hline
\end{tabular}
\caption{\label{table} Important packing fractions for $d=3,4,5$ and $6$.  
These include the equilibrium values for the freezing, $\phi_F$, melting, $\phi_M$,
and densest states, $\phi_{max}$, as well as the nonequilibrium MRJ values. The
freezing and melting points for $d=6$ were not calculated here.
$^*$Values computed in this work.}
\end{center}
\end{table*}

\begin{figure}
\begin{center}
\includegraphics*[width=0.55\columnwidth,keepaspectratio]{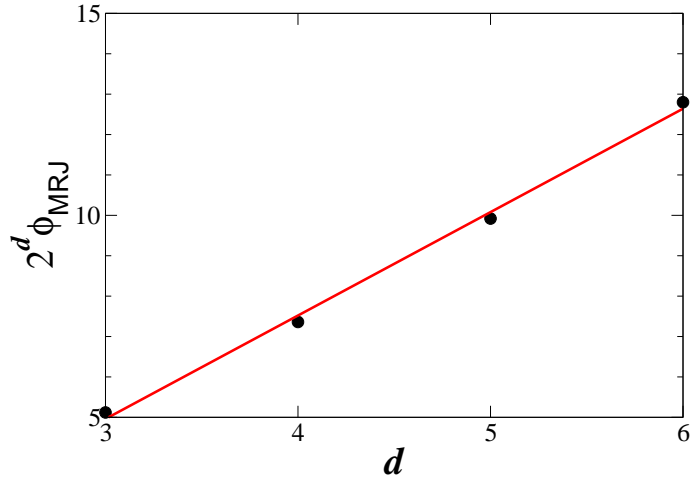}
\caption{(Color online)  Fit of the data for the product $2^d\phi_{MRJ} $ 
to the linear form  (\ref{linear}) for $3 \le d \le 6$ with $c_1=-2.72$ and $c_2=2.56$. }
\label{phi-fit}
\end{center}
\end{figure}

Our estimates for the MRJ packing fraction are compared 
to a theoretical formula proposed  by
Philipse~\cite{philipse} for the ``random jamming density'' $\phi_d$,
\begin{equation}
\phi_d \simeq \frac{0.046 d^2 + 1.22d + 0.73}{2^d}, \label{phil}
\end{equation}
which predicts $\phi_3 \simeq 0.601$, $\phi_4 \simeq 0.397$, $\phi_5
\simeq 0.249$, and $\phi_6 \simeq 0.152$. It is seen that Eq.~(\ref{phil})
underestimates MRJ density $\phi_{MRJ}$ in $d=3$ and becomes worse with increasing
dimension.  Following Ref. \cite{To06}, we obtain a better scaling
form by noting that the product $2^d \phi_{MRJ}$ for $3 \le d \le 6$ 
is well approximated by a function linear, rather than quadratic,  in $d$ (see Fig. \ref{phi-fit}), i.e., 
the scaling form for $\phi_{MRJ}$ is given by
\begin{equation}
\phi_{MRJ}= \frac{c_1}{2^d}+\frac{c_2 d}{2^d},
\label{linear}
\end{equation}
where $c_1=-2.72$ and $c_2=2.56$. Although the scaling form (\ref{linear}) applies only
in low dimensions such that $ d \ge 3$, theoretical arguments
given by Torquato, Uche and Stillinger \cite{To06} suggest
that the general scaling form (\ref{linear}) persists
in the high-dimensional asymptotic limit, albeit with different 
coefficients $c_1$ and $c_2$. In Ref. \cite{To06}, the density lower
bound $\phi_{MRJ} \ge (d+2)/2^d$ is derived for MRJ packings
in any dimension. This MRJ density lower bound yields
0.3125, 0.1875, 0.109375, 0.0625 for $d=3,4,5$ and 6, respectively.
We note that Parisi and Zamponi \cite{Pa06}
suggest the MRJ density scaling $\phi_{MRJ} \sim (d \log d) / 2^d$.

\subsection{Pair Correlations}

Our main interest is pair correlations in the jamming limit in four,
five and six dimensions.  We characterize jammed packings
statistically using the pair correlation function $g_2(r)$ and
structure factor $S(k)$.  The pair correlation function measures the
probability of finding a sphere center at a given distance from the
center of another sphere, normalized by the average number density
$\rho$ to go asymptotically to unity at large $r$; {\em i.e.}
\begin{equation}
g_2(r) = \frac{\langle P(r) \rangle}{\rho s_1(r)},
\end{equation}
where $P(r)$ is the probability density for finding a sphere center a
distance $r$ from an arbitrary sphere center, $\langle \rangle$
denotes an ensemble average, and $s_1(r)$ is the surface area of a
single hypersphere of radius $r$~\cite{torquatobook}: $s_1(r) = 2\pi^2
r^3$ in $d=4$, $s_1(r) = 8\pi^2 r^4/3$ in $d=5$ and $s_1(r) =\pi^3r^5$ in $d=6$.  
The structure factor
\begin{equation}
S(k) = 1 + \rho \hat h(k) \label{skdef}
\end{equation}
is related to the Fourier transform of the total correlation function
$h(r) = g_2(r)-1$. It measures spatial correlations at wavenumber $k$
and in particular, large-scale density fluctuations at
$k=0$~\cite{hyperuniformity}.  The structure factor can be observed
directly via scattering experiments~\cite{chaikin}.

\begin{figure}
\begin{center}
\includegraphics*[width=0.95\columnwidth,keepaspectratio]{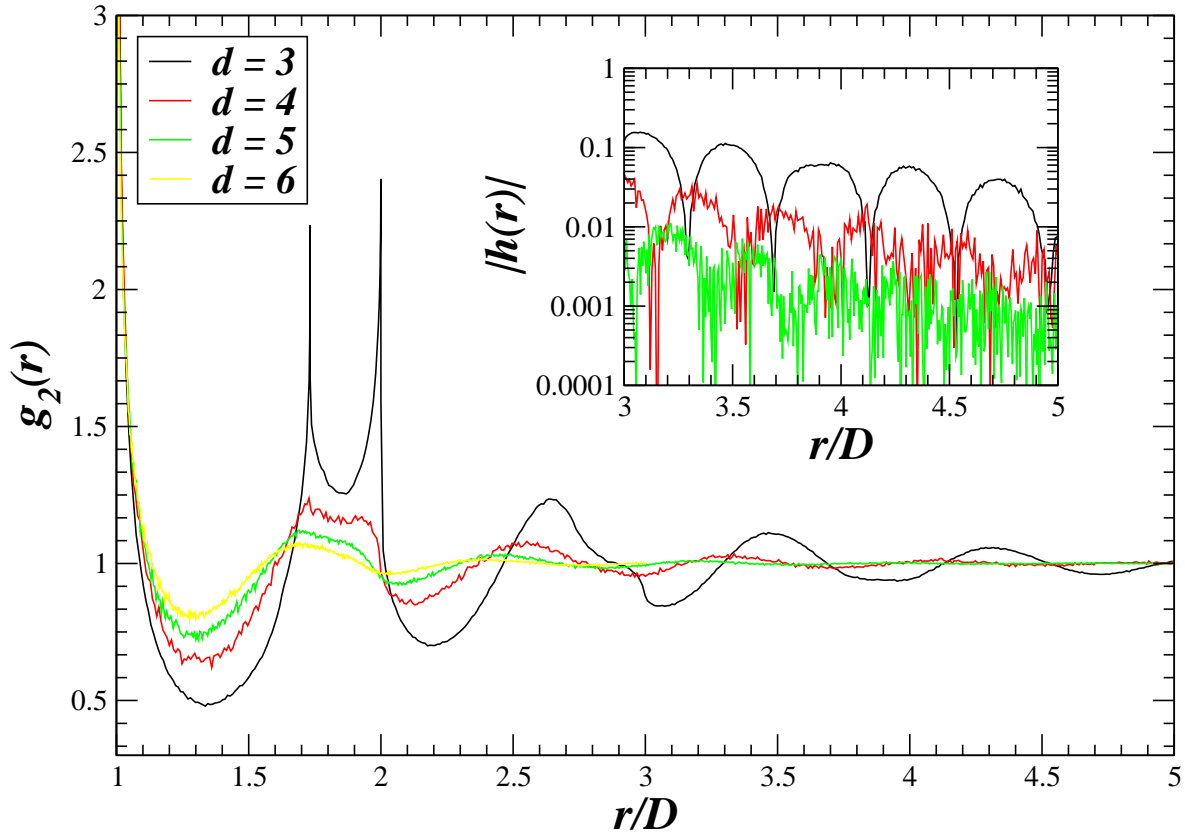}
\caption{The pair correlation function $g_2(r)$ for MRJ packings of
$10^4$ hard spheres for $d=3,~4,~5$ and $6$ at the respective
densities reported in Table \ref{table}. Pair separation is plotted in
units of the sphere diameter $D$.  [For $d=6$, $g_2(r)$ was only
calculated up to $r/D = 3$ due to the system size and periodic
boundary conditions].  The delta-function contribution [{\em cf.}
Eq.~\ref{delta}] at contact, of course, is not shown. The inset shows
$|h(r)| = |g_2(r)-1|$ on a logarithmic scale for $d=3,~4$ and
$5$. Each curve for $g_2(r)$ is obtained from a single packing
realization (not time-averaged). Curves for higher dimensions are
increasingly diminished.  }
\label{pcf}
\end{center}
\end{figure}

In the jamming limit, the pair correlation function $g_2(r)$ consists
of a $\delta$-function due to sphere contacts and a background part
$g_2^b(r)$ due to spheres not in contact:
\begin{equation}
g_2(r) = \frac{\bar Z \delta(r-D)}{\rho s_1(D)} + g_2^b(r),
\label{delta}
\end{equation}
where $\bar Z$ is the average kissing number.  Figure~\ref{pcf}
compares the pair correlation function for jammed packings of
$10^5$ spheres in $d=3$, $4$, $5$ and $6$.  Due to periodic
boundary conditions, $g_2(r)$ can only be calculated up to half the
length of the simulation box, which limits the calculation to $r/D
\simeq 3$ for $d=6$.  The well-known split second peak present in
$d=3$ is strongly diminished as the dimension increases, {\em i.e.,}
the amplitude of the split second peak decreases and the sharp cusps
become rounded with increasing dimension.  The split third peak
present in $d=3$ with considerable structure and two shoulders
vanishes almost completely in the higher dimensions. The oscillations
are strongly damped with increasing dimension and the period of
oscillations might also decrease slightly with increasing dimension;
this latter possibility is revealed more vividly in the structure
factor through the shift in the location of the maximum, as we will
describe below.  The inset to Fig.~\ref{pcf} shows the magnitude of
the decaying oscillations in $h(r)$ on a semi-log scale.  Though at
the values of $r/D$ shown, up to about half the length of the
simulation box, there is still structure in addition to the
oscillations, especially apparent for $d=3$, it appears that the decay
rate of the oscillations in $h(r)$ does not change significantly with
dimension, whereas the amplitude of oscillations does.  However,
further studies with larger $r$ and therefore larger systems are
needed to obtain more quantitative results.

We calculate the structure factor $S(k)$, defined in Eq.~\ref{skdef},
for $d=4$ and $d=5$ by
\begin{equation}
 S(K) = 1 + 128\phi \int_0^{\infty} x^3 h(x) \frac{J_1(Kx)}{Kx}dx
 \label{s4D}
\end{equation}
and 
\begin{equation}
S(K) = 1 + 480 \phi \int_0^{\infty} \frac{x^4 h(x)}{(Kx)^2}
 \left[\frac{\sin(Kx)}{Kx} - \cos(Kx) \right] dx, \label{s5D}
\end{equation}
respectively, where $\phi = \pi^2 \rho D^4/32$ for $d=4$ and $\phi =
\pi^2 \rho D^5/60$ for $d=5$, $x=r/D$ and $K=kD$ are the dimensionless
radius and wave number, and $J_{\nu}(x)$ is the Bessel function of
order $\nu$.  We do not calculate the structure factor for $d=6$
because at present we do not have $g_2(r)$ over a sufficiently large
range of $r$.

Following Ref.~\cite{alekshyper}, rather than working directly with
$g_2(x)$ as in Eq.~(\ref{skdef}), we consider the average cumulative
coordination $Z(x)$, defined to be the following volume integral of
$g_2(x)$:  
\begin{equation}
Z(x) = \rho \int_1^x s_1(x') g_2(x') dx'. \label{Zintegralg}
\end{equation}
The excess coordination $\Delta Z(x)$,
\begin{eqnarray}
\Delta Z(x) & = & 1 + 64 \phi \int_0^x (x')^3 h(x') dx' \\
\Delta Z(x) & = & 1 + 160 \phi \int_0^x (x')^4 h(x') dx',
\end{eqnarray}
for $d=4$ and $d=5$, respectively, is the average excess number of
sphere centers inside a spherical window of radius $x$ centered at a
sphere, compared to the ideal gas expectations, $16 \phi x^4$ for
$d=4$ and $32 \phi x^5$ in $d=5$.  We can rewrite Eq.~(\ref{skdef}) in
terms of $\Delta Z(x)$ using integration by parts to get
\begin{equation}
S(K) = -
2 \int_0^{\infty} \Delta Z(x) \frac{d}{dx}\frac{J_1(K x)}{K x} dx
\end{equation}
and
\begin{equation}
S(K) = -3\int_0^{\infty}
\Delta Z(x) \frac{d}{dx}\left[\frac{\sin(K x)}{(K x)^3} -
\frac{\cos(K x)}{(K x)^2} \right] dx,
\end{equation}
for $d=4$ and $d=5$, respectively.  Note that accurate evaluations
of the integrals of $\Delta Z(x)$ require extrapolations of its
large-$x$ tail behavior, for which we have used an exponentially-damped 
oscillating function \cite{footnoteZ}.

\begin{figure}
\begin{center}
\includegraphics*[width=0.95\columnwidth,keepaspectratio]{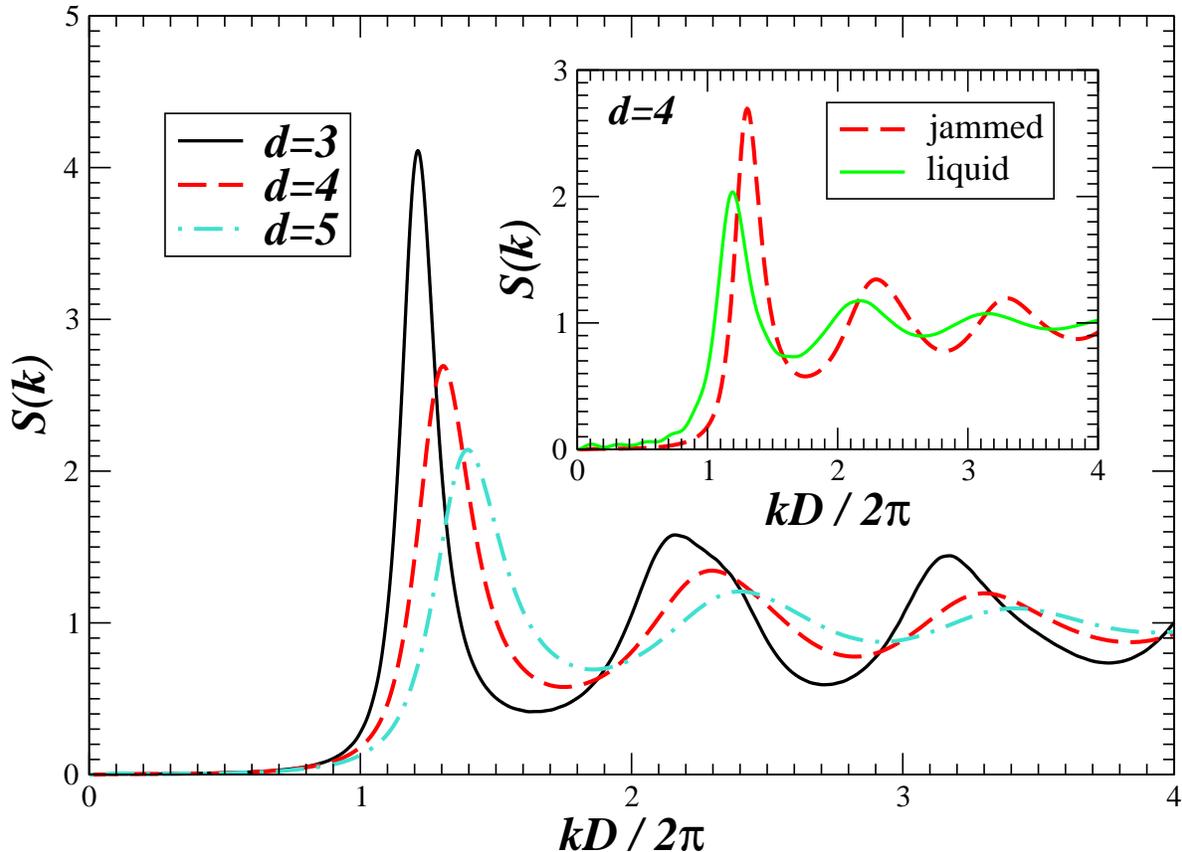}
\caption{The structure factor $S(k)$ for jammed packings of $10^5$
spheres for $d=3$, $4$ and $5$ at the respective densities reported in
Table \ref{table}. Inset: A comparison for $d=4$ of $S(k)$ for a
jammed packing and for a fluid near the freezing point ($\phi\approx
0.31$). Each curve for $S(k)$ is obtained from a single packing
realization (not time averaged). }
\label{sk}
\end{center}
\end{figure}

Figure~\ref{sk} shows $S(k)$ for jammed packings of $10^5$ spheres in
three, four and five dimensions.  Qualitatively, $S(k)$ is somewhat
similar for $d=3$, $4$, and $5$.  However, with increasing dimension,
the height of the first peak of $S(k)$ decreases, the location of the
first peak moves to smaller wavelengths, and the oscillations become
damped.  The width of the first peak also increases with increasing
dimension, which could indicate that the correlation length decreases
with increasing dimension.  The inset to Fig.~\ref{sk} shows $S(k)$
for a jammed packing and a fluid near the freezing point in four
dimensions.  The relation between the structure factor for the fluid
and jammed packing is strikingly similar to what is found for $d=3$,
except that the peaks of both curves for $d=4$ appear scaled down
relative to $d=3$.  Overall, our results for both $g_2(r)$ and $S(k)$
are consistent with a recently proposed ``decorrelation''
principle~\cite{decorrelation}. We note that similar pair
decorrelations are observed for RSA packings as the dimension
increases up to $d=6$ \cite{To06}.

It is of interest to determine whether infinite-wavelength density fluctuations
$S(k=0)$ vanish; systems with this property are called
``hyperuniform''~\cite{hyperuniformity}.  For equilibrium fluids and
crystals, $S(k = 0)$ is proportional to the isothermal compressibility
and therefore must be positive.  As for $d=3$, $S(k)$ for $d=4$
appears to go to zero faster near the origin for the jammed packing
than for the fluid.  However, we cannot reliably determine whether
$S(k)$ vanishes at the origin because our calculation of $S(k)$ for
small $k$ involved an extrapolation of the large-$x$ tail of $\Delta
Z(x)$.  Nevertheless, using larger system sizes of one million
spheres, saturated~\cite{saturated} MRJ packings for $d=3$ have been
shown to be hyperuniform to a high accuracy~\cite{alekshyper} and the
comparison of $d=4$ and $d=5$ to $d=3$, shown in Fig.~\ref{sk},
suggests that MRJ packings for $d=4$ and $d=5$ are also hyperuniform.

\subsection{Isostaticity}

\begin{figure}
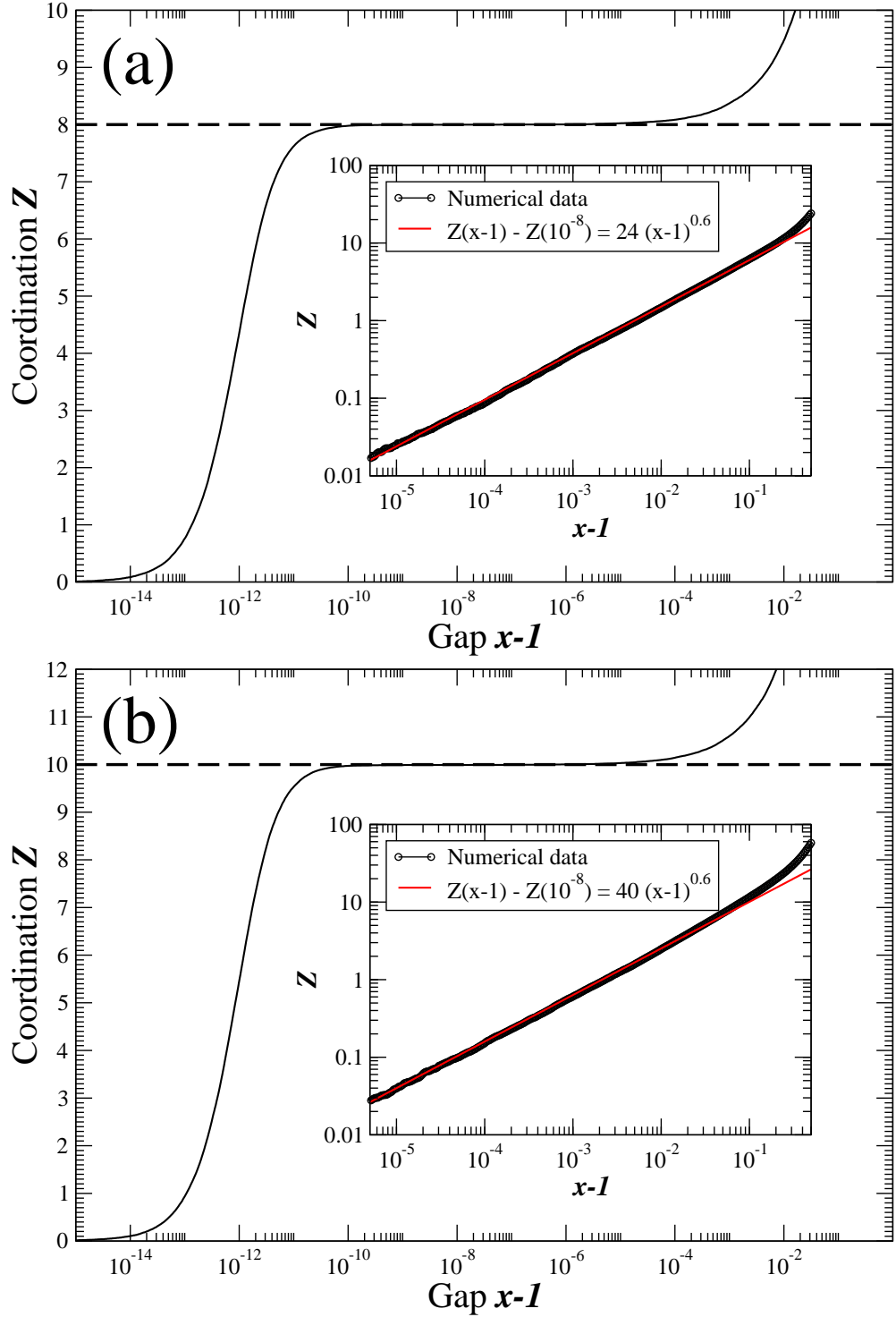

\begin{center}
\includegraphics*[width=0.8\columnwidth,keepaspectratio]{logzwor4D.eps}
\includegraphics*[width=0.8\columnwidth,keepaspectratio]{logzpower5D.eps}
\caption{The near-contact cumulative coordination $Z(x)$ [{\em c.f.}
Eq.~\ref{Zintegralg}] for $10^4$-sphere MRJ packings for $d=4$ (a)
and for $d=5$ (b), with rattlers removed.  The inset shows $Z(x)$ on a
log-log scale along with power-law fits for intermediate interparticle
gap $x-1$ beyond contact. $10^5$-sphere MRJ packings in $d=5$ with
final expansion rates of $\gamma = 10^{-4}$ give similar results; such
packings with final expansion rates of $\gamma = 10^{-5}$ are
(presently) too computationally expensive.  Compare these plots to the
equivalent results for $d=3$ in Ref.~\cite{aleks}~[{\em c.f.}
Fig.~8].  }
\label{logzpower}
\end{center}
\end{figure}

We study the near-contact contribution to $g_2(r)$, {\em i.e.,}
interparticle distances $r$ that are very close to the sphere diameter
$D$, using the cumulative coordination number $Z(x)$, where as before
$x=r/D$ is the dimensionless radius and $x-1$ is the dimensionless
interparticle gap.  Figure~\ref{logzpower} shows $Z(x)$ for jammed
packings of $10,000$ spheres for $d=4$ and $d=5$ with rattlers
removed~\cite{no6Disostaticity}.  The plateaus at $Z=8$ in
Fig.~\ref{logzpower} (a) and $Z=10$ in Fig.~\ref{logzpower} (b) show
that both packings are isostatic.  Isostatic packings are jammed
packings which have the minimal number of contacts necessary for
collective jamming.  For spheres, this occurs when the number of
degrees of freedom equals the number of contacts (or constraints);
each $d$-dimensional sphere has $d$ degrees of freedom, and hence the
mean number of contacts experienced by a sphere necessary for jamming
is $2d$, since each contact involves two spheres.

Packings produced by the LS algorithm almost always contain a nonzero
fraction of ``rattlers'', which are spheres trapped in a cage of
jammed neighbors, but free to move within the cage.  We find
approximately $\sim 1\%$ rattlers for $d=4$ and $\sim 0.6\%$ rattlers
for $d=5$, as compared to $\sim 2-3\%$ rattlers for
$d=3$~\cite{aleks}.  Rattlers can be identified as having less than
the required $d+1$ contacts necessary for local jamming and are
removed to study the jammed backbone of the packing, which we focus on
in this section.

The insets to Fig.~\ref{logzpower} (a) and (b) show $Z(x) - 2d$, along
with a power-law fit for intermediate interparticle gap $x-1$,
\begin{equation}
Z(x) = \bar Z + Z_0 (x-1)^{\alpha},
\end{equation}
where $\bar Z = 2d$.  Since the packings are generally slightly
subisostatic, we apply a small correction ($<0.1\%$) to the isostatic
prediction of $2d$ by using the midpoint of the apparent plateau in
$Z(x)$.  The best-fit exponent is $\alpha \simeq 0.6$ in both $d=4$
and $d=5$, in agreement with that found for $d=3$~\cite{aleks}.  The
coefficients of the power law, $Z_0 \simeq 11$ in $d=3$, $Z_0 \simeq
24$ for $d=4$, and $Z_0 \simeq 40$ for $d=5$ are close to the
corresponding kissing numbers of the densest packings, $Z = 12$ for
$d=3$, $Z = 24$ for $d=4$, $40 \leq Z \leq 46$ for $d=5$ and $72 \leq
Z \leq 80$ for $d=6$.  Motivated by this observation, we measured the
value of the gap $x-1$ at which the cumulative coordination $Z(x)$
equals the kissing number of the densest packing to be: $x-1 \simeq
0.35$, $0.34$, $0.31-0.36$ and $0.33-0.36$ in $d=3$, $4$, $5$ and $6$,
respectively, which we can define to be the cutoff for the
near-neighbor shell.  This definition produces results similar to that
of the more common definition of the cutoff for the near-neighbor
shell as the value of the gap $x-1$ at the first minimum in $g_2$,
which occurs at $x-1 \simeq 0.35$, $0.32$, $0.30$ and $0.28$ in $d=3$,
$4$, $5$ and $6$, respectively.  It is also interesting to observe
that the power-law fit to $Z(x)$ is good over a rather wide range of
gaps, almost up to the first minimum in $g_2$. We should, however,
emphasize that the minimum of $g_2$ is not very precisely defined,
especially due to decorrelation in high dimensions, and the choice of
the gap at the minimum of $g_2$, or at which $Z(x)$ equals the kissing
number of the densest packing, as a special point is somewhat
arbitrary and not theoretically justified at present.

\section{Discussion}

We have presented the first numerical results characterizing random
jammed hard-sphere packings in four, five and six dimensions.  We find
disordered packings, representative of the maximally random jammed
state, to be isostatic and have packing fractions $\phi_{MRJ} \simeq
0.46$, $\phi_{MRJ} \simeq 0.31$ and $\phi_{MRJ} \simeq 0.20$ for
$d=4$, $5$ and $6$, respectively.  For equilibrium sphere packings, we
estimate the freezing and melting packing fractions for the
fluid-solid transition in four dimensions to be $\phi_F \simeq 0.32$
and $\phi_M \simeq 0.39$, respectively, and in five dimensions to be
$\phi_F \simeq 0.19$ and $\phi_M \simeq 0.24$, respectively.
Additionally, a signature characteristic of the kinetic glass
transition is observed around $\phi_G \simeq 0.41$ for $d=4$.  We
observe a significantly lower tendency to crystallize for $d=4$ than
in $d=3$, which is likely due to the closer proximity of the melting
and kinetic glass transition densities for $d=4$~\cite{binarydisk}.

We find that in high dimensions the split-second peak in the
pair correlation function $g_2$, present for $d=3$, gets dramatically
diminished and oscillations in both $g_2$ and the structure factor
$S(k)$ get significantly dampened.  These findings are consistent with
a recently proposed ``decorrelation principle''~\cite{decorrelation},
stating that unconstrained spatial correlations vanish asymptotically
in the high-dimensional limit and that the
$n$-particle correlation function $g_n$ for any $n \ge 3$ can be 
inferred entirely from a knowledge of the number density $\rho$ and the pair 
correlation function $g_2({\bf r})$.  Accordingly, in this limit the
pair correlation function $g_2(r)$ would be expected to retain the
delta-function contribution from nearest-neighbor contacts, but the
extra structure representing unconstrained spatial correlations beyond
a single sphere diameter would vanish.  Figures~\ref{pcf} and~\ref{sk}
show dramatically the decorrelation principle already taking effect in
four, five and six dimensions.  We note that decorrelation principle
is also apparent in the same dimensions for RSA packings \cite{To06}.

One should not be misled to believe that the decorrelation principle
is an expected  ``mean-field" behavior.  For example, it
is well known that in some spin systems correlations vanish 
in the limit $d \rightarrow \infty$ and the system approaches the mean-field
behavior. While this idea has meaning for spin
systems with attractive interactions, hard-core systems,
whose total potential energy is either zero or infinite, cannot be characterized by a mean field.
Mean-field theories are limited
to equilibrium considerations, and thus do not distinguish between ``constrained" and ``unconstrained" correlations 
because, unlike us, they are not concerned with non-equilibrium packings of which there 
an infinite number of distinct ensembles. The decorrelation
principle is a statement about any disordered packing, equilibrium or not. For example, contact delta functions are
an important attribute of non-equilibrium jammed disordered packings and have no analog in equilibrium lattice
models of any dimension. The decorrelation principle is also justified
on the basis of a rigorous upper bound on the maximal packing density in
high dimensions \cite{decorrelation},
which has no counterpart in mean-field theories.

A particularly interesting property of jammed hard-sphere packings is
hyperuniformity, the complete suppression of infinite wavelength
density fluctuations, {\em i.e.,} the vanishing of the structure
factor $S(k)$ as $k\rightarrow 0$.  It has been recently conjectured
that all saturated strictly-jammed packings are
hyperuniform~\cite{hyperuniformity} and calculations of the structure
factor near $k = 0$ for $d=3$ using one million particle systems have
strongly suggested that MRJ packings for $d=3$ are indeed
hyperuniform~\cite{alekshyper}.  Though the system sizes used in this
paper were too small to probe such large-scale density fluctuations
without relying on dubious extrapolations, our numerical results for
the structure factor for $d=4$ and $d=5$, as shown in Fig.~\ref{sk},
are consistent with hyperuniformity.

As in three dimensions, disordered jammed sphere packings show no
signs of crystallization, are isostatic, and have a power-law
divergence in $g_{2}(r)$ at contact.  Interestingly, all three
dimensions ($3$, $4$ and $5$) share the same power law exponent
$1-\alpha \simeq 0.4$ when rattlers are removed, and show the first minimum
of $g_{2}(r)$ close to where the cumulative coordination $Z(r)$ equals
the kissing number of the densest lattice packing.  Such a relation
between the kissing numbers of the densest packings and MRJ packings
for $d=3$, $4$, $5$ and $6$, if not coincidental, is very surprising
and may be a consequence of the geometrical structure of MRJ packings.
It suggests that disordered packings might be deformed crystal
packings, in which the true contacts are deformed into near contacts,
and only the minimal number of contacts necessary for jamming is
preserved. This interpretation is to be contrasted with the usual
interpretation of disordered packing in $d=3$ in terms of tetrahedral
or icosahedral packings, without relation to the crystal (FCC)
packing. The former interpretation is similar to the one of the MRJ state 
for binary hard disks as a random partitioning of the monodisperse 
triangular crystal into ``small'' and ``large'' disks, i.e., a deformed 
monodisperse triangular disk crystal in which a randomly chosen 
fraction of the particles have grown in size, as proposed in Ref. 
\cite{binarydisk}.

It is important to point out that hard-sphere packings behave rather
differently in two dimensions than in three and higher dimensions.
For $d=2$, jammed hard-sphere systems are polycrystalline and there is
a very weak, nearly continuous fluid-solid phase transition.  Hence,
there is no glassy behavior for $d=2$ and consequently no amorphous
jammed packings.  Glassy behavior, due to geometrical frustration
arising from the inconsistency of local optimal packing rules and
global packing constraints, first appears in three
dimensions~\cite{torquatobook}.  It is likely that geometrical
frustration generally increases with dimension, consistent with our observation
that nucleation is suppressed with increasing dimension.

Computational costs rise dramatically with increasing dimension and
theoretical understanding based on observations in moderate dimensions
is necessary.  We believe that the numerical results presented in this
work provide tests and motivations for such theories.

\acknowledgments M. S. was supported by the National Science
Foundation and A. D. and S. T. were partially supported by the
National Science Foundation under Grant No. DMS-0312067.

%\bibliography{sphereref}

\begin{thebibliography}{44}
\expandafter\ifx\csname natexlab\endcsname\relax\def\natexlab#1{#1}\fi
\expandafter\ifx\csname bibnamefont\endcsname\relax
  \def\bibnamefont#1{#1}\fi
\expandafter\ifx\csname bibfnamefont\endcsname\relax
  \def\bibfnamefont#1{#1}\fi
\expandafter\ifx\csname citenamefont\endcsname\relax
  \def\citenamefont#1{#1}\fi
\expandafter\ifx\csname url\endcsname\relax
  \def\url#1{\texttt{#1}}\fi
\expandafter\ifx\csname urlprefix\endcsname\relax\def\urlprefix{URL }\fi
\providecommand{\bibinfo}[2]{#2}
\providecommand{\eprint}[2][]{\url{#2}}

\bibitem[{\citenamefont{{Torquato} et~al.}(2000)\citenamefont{{Torquato},
  {Truskett}, and {Debenedetti}}}]{rcp}
\bibinfo{author}{\bibfnamefont{S.}~\bibnamefont{{Torquato}}},
  \bibinfo{author}{\bibfnamefont{T.~M.} \bibnamefont{{Truskett}}},
  \bibnamefont{and} \bibinfo{author}{\bibfnamefont{P.~G.}
  \bibnamefont{{Debenedetti}}}, \bibinfo{journal}{Phys. Rev. Lett.}
  \textbf{\bibinfo{volume}{84}}, \bibinfo{pages}{2064} (\bibinfo{year}{2000}).

\bibitem[{\citenamefont{{Kansal} et~al.}(2002)\citenamefont{{Kansal},
  {Torquato}, and {Stillinger}}}]{kansal}
\bibinfo{author}{\bibfnamefont{A.~R.} \bibnamefont{{Kansal}}},
  \bibinfo{author}{\bibfnamefont{S.}~\bibnamefont{{Torquato}}},
  \bibnamefont{and} \bibinfo{author}{\bibfnamefont{F.~H.}
  \bibnamefont{{Stillinger}}}, \bibinfo{journal}{Phys. Rev. E}
  \textbf{\bibinfo{volume}{66}}, \bibinfo{pages}{041109}
  (\bibinfo{year}{2002}).

\bibitem[{\citenamefont{{Donev}
  et~al.}(2005{\natexlab{a}})\citenamefont{{Donev}, {Torquato}, and
  {Stillinger}}}]{aleks}
\bibinfo{author}{\bibfnamefont{A.}~\bibnamefont{{Donev}}},
  \bibinfo{author}{\bibfnamefont{S.}~\bibnamefont{{Torquato}}},
  \bibnamefont{and} \bibinfo{author}{\bibfnamefont{F.~H.}
  \bibnamefont{{Stillinger}}}, \bibinfo{journal}{Phys. Rev. E}
  \textbf{\bibinfo{volume}{71}}, \bibinfo{pages}{011105}
  (\bibinfo{year}{2005}{\natexlab{a}}).

\bibitem[{\citenamefont{{Donev}
  et~al.}(2005{\natexlab{b}})\citenamefont{{Donev}, {Stillinger}, and
  {Torquato}}}]{alekshyper}
\bibinfo{author}{\bibfnamefont{A.}~\bibnamefont{{Donev}}},
  \bibinfo{author}{\bibfnamefont{F.~H.} \bibnamefont{{Stillinger}}},
  \bibnamefont{and}
  \bibinfo{author}{\bibfnamefont{S.}~\bibnamefont{{Torquato}}},
  \bibinfo{journal}{Phys. Rev. Lett.} \textbf{\bibinfo{volume}{95}},
  \bibinfo{pages}{090604} (\bibinfo{year}{2005}{\natexlab{b}}).

\bibitem[{\citenamefont{{Rintoul} and {Torquato}}(1996)}]{rintoul}
\bibinfo{author}{\bibfnamefont{M.~D.} \bibnamefont{{Rintoul}}}
  \bibnamefont{and}
  \bibinfo{author}{\bibfnamefont{S.}~\bibnamefont{{Torquato}}},
  \bibinfo{journal}{Phys. Rev. Lett.} \textbf{\bibinfo{volume}{77}},
  \bibinfo{pages}{4198} (\bibinfo{year}{1996}).

\bibitem[{\citenamefont{{Donev} et~al.}(submitted)\citenamefont{{Donev},
  {Stillinger}, and {Torquato}}}]{binarydisk}
\bibinfo{author}{\bibfnamefont{A.}~\bibnamefont{{Donev}}},
  \bibinfo{author}{\bibfnamefont{F.~H.} \bibnamefont{{Stillinger}}},
  \bibnamefont{and}
  \bibinfo{author}{\bibfnamefont{S.}~\bibnamefont{{Torquato}}},
   \bibinfo{journal}{Phys. Rev. Lett.} \textbf{\bibinfo{volume}{96}},
     \bibinfo{pages}{225502} (\bibinfo{year}{2006}).


\bibitem[{\citenamefont{{Finney}}(1970)}]{finney}
\bibinfo{author}{\bibfnamefont{J.~L.} \bibnamefont{{Finney}}},
  \bibinfo{journal}{Proc. R. Soc. Lond. A} \textbf{\bibinfo{volume}{319}},
  \bibinfo{pages}{479} (\bibinfo{year}{1970}).

\bibitem[{\citenamefont{Bennett}(1972)}]{bennett}
\bibinfo{author}{\bibfnamefont{C.~H.} \bibnamefont{Bennett}},
  \bibinfo{journal}{J. Appl. Phys.} \textbf{\bibinfo{volume}{32}},
  \bibinfo{pages}{2727} (\bibinfo{year}{1972}).

\bibitem[{\citenamefont{{Tobochnik} and {Chapin}}(1988)}]{tobochnik}
\bibinfo{author}{\bibfnamefont{J.}~\bibnamefont{{Tobochnik}}} \bibnamefont{and}
  \bibinfo{author}{\bibfnamefont{P.~M.} \bibnamefont{{Chapin}}},
  \bibinfo{journal}{J. Chem. Phys.} \textbf{\bibinfo{volume}{88}},
  \bibinfo{pages}{5824} (\bibinfo{year}{1988}).

\bibitem[{\citenamefont{{Zinchenko}}(1994)}]{zinchenko}
\bibinfo{author}{\bibfnamefont{A.~Z.} \bibnamefont{{Zinchenko}}},
  \bibinfo{journal}{J. Comput. Phys.} \textbf{\bibinfo{volume}{114}},
  \bibinfo{pages}{298} (\bibinfo{year}{1994}).

\bibitem[{\citenamefont{Xu et~al.}(2005)\citenamefont{Xu, Blawzdziewicz, and
  O'Hern}}]{ohern}
\bibinfo{author}{\bibfnamefont{N.}~\bibnamefont{Xu}},
  \bibinfo{author}{\bibfnamefont{J.}~\bibnamefont{Blawzdziewicz}},
  \bibnamefont{and} \bibinfo{author}{\bibfnamefont{C.}~\bibnamefont{O'Hern}},
  \bibinfo{journal}{Phys. Rev. E} \textbf{\bibinfo{volume}{71}},
  \bibinfo{pages}{061306} (\bibinfo{year}{2005}).

\bibitem[{\citenamefont{{Chaikin} and {Lubensky}}(1995)}]{chaikin}
\bibinfo{author}{\bibfnamefont{P.~M.} \bibnamefont{{Chaikin}}}
  \bibnamefont{and} \bibinfo{author}{\bibfnamefont{T.~C.}
  \bibnamefont{{Lubensky}}}, \emph{\bibinfo{title}{{Principles of condensed
  matter physics}}} (\bibinfo{publisher}{Cambridge University Press},
  \bibinfo{address}{Cambridge, United Kingdom}, \bibinfo{year}{1995}).

\bibitem[{\citenamefont{{Luban} and {Baram}}(1982)}]{virial234}
\bibinfo{author}{\bibfnamefont{M.}~\bibnamefont{{Luban}}} \bibnamefont{and}
  \bibinfo{author}{\bibfnamefont{A.}~\bibnamefont{{Baram}}},
  \bibinfo{journal}{J. Chem. Phys.} \textbf{\bibinfo{volume}{76}},
  \bibinfo{pages}{3233} (\bibinfo{year}{1982}).

\bibitem[{\citenamefont{{Joslin}}(1982)}]{virial234too}
\bibinfo{author}{\bibfnamefont{C.~J.} \bibnamefont{{Joslin}}},
  \bibinfo{journal}{J. Chem. Phys.} \textbf{\bibinfo{volume}{77}},
  \bibinfo{pages}{2701} (\bibinfo{year}{1982}).

\bibitem[{\citenamefont{{Luban} and {Michels}}(1990)}]{EOSlubanmichels}
\bibinfo{author}{\bibfnamefont{M.}~\bibnamefont{{Luban}}} \bibnamefont{and}
  \bibinfo{author}{\bibfnamefont{J.~P.~J.} \bibnamefont{{Michels}}},
  \bibinfo{journal}{Phys. Rev. A} \textbf{\bibinfo{volume}{41}},
  \bibinfo{pages}{6796} (\bibinfo{year}{1990}).

\bibitem[{\citenamefont{{Bishop} et~al.}(1999)\citenamefont{{Bishop},
  {Masters}, and {Clarke}}}]{virial56}
\bibinfo{author}{\bibfnamefont{M.}~\bibnamefont{{Bishop}}},
  \bibinfo{author}{\bibfnamefont{A.}~\bibnamefont{{Masters}}},
  \bibnamefont{and} \bibinfo{author}{\bibfnamefont{J.~H.~R.}
  \bibnamefont{{Clarke}}}, \bibinfo{journal}{J. Chem. Phys.}
  \textbf{\bibinfo{volume}{110}}, \bibinfo{pages}{11449}
  (\bibinfo{year}{1999}); \bibinfo{author}{\bibfnamefont{M.}~\bibnamefont{{Bishop}}},
  \bibinfo{author}{\bibfnamefont{A.}~\bibnamefont{{Masters}}},
  \bibnamefont{and} \bibinfo{author}{\bibfnamefont{A.~Y.}
  \bibnamefont{{Vlasov}}}, \bibinfo{journal}{J. Chem. Phys.}
  \textbf{\bibinfo{volume}{121}}, \bibinfo{pages}{6884} (\bibinfo{year}{2004});
  M. Bishop, A. Masters and A. Yu. Vlasov, J. Chem. Phys. {\bf 122},
    154502 (2005); M. Bishop and P.A. Whitlock, J. Chem. Phys. {\bf 123}, 014507 (2005).



\bibitem[{\citenamefont{Clisby and McCoy}(2004)}]{Virial4_EvenDimensions}
\bibinfo{author}{\bibfnamefont{N.}~\bibnamefont{Clisby}} \bibnamefont{and}
  \bibinfo{author}{\bibfnamefont{B.}~\bibnamefont{McCoy}}, \bibinfo{journal}{J.
  Stat. Phys.} \textbf{\bibinfo{volume}{114}}, \bibinfo{pages}{1343}
  (\bibinfo{year}{2004}); ibid., 1361 (2004); ibid., {\bf 122}, 15 (2006);
N. Clisby and B.M. McCoy, J. Phys. {\bf 64}, 775 (2005).  

\bibitem{Lyberg} I. Lyberg, J. Stat. Phys. {\bf 119}, 747 (2005).

\bibitem[{\citenamefont{Clisby and McCoy}(2006)}]{Virial89_HighDimensions}
\bibinfo{author}{\bibfnamefont{N.}~\bibnamefont{Clisby}} \bibnamefont{and}
  \bibinfo{author}{\bibfnamefont{B.}~\bibnamefont{McCoy}}, \bibinfo{journal}{J.
  Stat. Phys.} \textbf{\bibinfo{volume}{122}}, \bibinfo{pages}{15}
  (\bibinfo{year}{2006}).



\bibitem[{\citenamefont{Frisch and Percus}(1999)}]{frisch}
\bibinfo{author}{\bibfnamefont{H.~L.} \bibnamefont{Frisch}} \bibnamefont{and}
  \bibinfo{author}{\bibfnamefont{J.~K.} \bibnamefont{Percus}},
  \bibinfo{journal}{Phys. Rev. E} \textbf{\bibinfo{volume}{60}},
  \bibinfo{pages}{2942} (\bibinfo{year}{1999}).

\bibitem[{\citenamefont{Parisi and Slanina}(2000)}]{parisi}
\bibinfo{author}{\bibfnamefont{G.}~\bibnamefont{Parisi}} \bibnamefont{and}
  \bibinfo{author}{\bibfnamefont{F.}~\bibnamefont{Slanina}},
  \bibinfo{journal}{Phys. Rev. E} \textbf{\bibinfo{volume}{62}},
  \bibinfo{pages}{6554} (\bibinfo{year}{2000}).


\bibitem[{\citenamefont{{Finken} et~al.}(2001)\citenamefont{{Finken},
  {Schmidt}, and {Lowen}}}]{finken}
\bibinfo{author}{\bibfnamefont{R.}~\bibnamefont{{Finken}}},
  \bibinfo{author}{\bibfnamefont{M.}~\bibnamefont{{Schmidt}}},
  \bibnamefont{and} \bibinfo{author}{\bibfnamefont{H.}~\bibnamefont{{Lowen}}},
  \bibinfo{journal}{Phys. Rev. E} \textbf{\bibinfo{volume}{65}},
  \bibinfo{pages}{016108} (\bibinfo{year}{2001}).


\bibitem[{\citenamefont{{Torquato}}(2002)}]{torquatobook}
\bibinfo{author}{\bibfnamefont{S.}~\bibnamefont{{Torquato}}},
  \emph{\bibinfo{title}{{Random Heterogeneous Materials: Microstructure and
  Macroscopic Properties}}} (\bibinfo{publisher}{Springer-Verlag},
  \bibinfo{address}{New York}, \bibinfo{year}{2002}).

\bibitem[{\citenamefont{{Philipse}}(2003)}]{philipse}
\bibinfo{author}{\bibfnamefont{A.~P.} \bibnamefont{{Philipse}}},
  \bibinfo{journal}{Colloids and Surfaces A} \textbf{\bibinfo{volume}{213}},
  \bibinfo{pages}{167} (\bibinfo{year}{2003}).

\bibitem[{\citenamefont{{Torquato} and {Stillinger}}(2003)}]{hyperuniformity}
\bibinfo{author}{\bibfnamefont{S.}~\bibnamefont{{Torquato}}} \bibnamefont{and}
  \bibinfo{author}{\bibfnamefont{F.~H.} \bibnamefont{{Stillinger}}},
  \bibinfo{journal}{Phys. Rev. E} \textbf{\bibinfo{volume}{68}},
  \bibinfo{pages}{041113} (\bibinfo{year}{2003}).


\bibitem[{\citenamefont{Parisi and Zamponi}(2006)}]{Pa06}
\bibinfo{author}{\bibfnamefont{G.}~\bibnamefont{Parisi}} \bibnamefont{and}
  \bibinfo{author}{\bibfnamefont{F.}~\bibnamefont{Zamponi}},
  \bibinfo{journal}{J. Stat. Mech.} p. \bibinfo{pages}{P03017}
  (\bibinfo{year}{2006}).



\bibitem[{\citenamefont{{Torquato} and {Stillinger}}(in press)}]{decorrelation}
\bibinfo{author}{\bibfnamefont{S.}~\bibnamefont{{Torquato}}} \bibnamefont{and}
  \bibinfo{author}{\bibfnamefont{F.~H.} \bibnamefont{{Stillinger}}},
  \bibinfo{journal}{Experimental Math.}  (\bibinfo{year}{in press}).
  
  

\bibitem[{\citenamefont{{Torquato} and {Stillinger}}(2006)}]{torquato15}
\bibinfo{author}{\bibfnamefont{S.}~\bibnamefont{{Torquato}}} \bibnamefont{and}
  \bibinfo{author}{\bibfnamefont{F.~H.} \bibnamefont{{Stillinger}}},
  \bibinfo{journal}{Phys. Rev. E} \textbf{\bibinfo{volume}{73}},
  \bibinfo{pages}{031106} (\bibinfo{year}{2006}).
  

\bibitem{To06}
S. Torquato, O. U. Uche and F. H. Stillinger, in preparation.


\bibitem[{\citenamefont{{Conway} and {Sloane}}(1998)}]{conway}
\bibinfo{author}{\bibfnamefont{J.~H.} \bibnamefont{{Conway}}} \bibnamefont{and}
  \bibinfo{author}{\bibfnamefont{N.~J.~A.} \bibnamefont{{Sloane}}},
  \emph{\bibinfo{title}{{Sphere Packings, Lattices, and Groups}}}
  (\bibinfo{publisher}{Springer-Verlag}, \bibinfo{address}{New York},
  \bibinfo{year}{1998}).


\bibitem{Lue}
L. Lue, J. Chem. Phys. {\bf 122}, 044513 (2005).


\bibitem[{\citenamefont{{Bishop} et~al.}(2005)\citenamefont{{Bishop},
  {Whitlock}, and {Klein}}}]{bishop}
\bibinfo{author}{\bibfnamefont{M.}~\bibnamefont{{Bishop}}},
  \bibinfo{author}{\bibfnamefont{P.}~\bibnamefont{{Whitlock}}},
  \bibnamefont{and} \bibinfo{author}{\bibfnamefont{D.}~\bibnamefont{{Klein}}},
  \bibinfo{journal}{J. Chem. Phys.} \textbf{\bibinfo{volume}{122}},
  \bibinfo{pages}{074508} (\bibinfo{year}{2005}).


\bibitem[{\citenamefont{{Michels} and {Trappaniers}}(1984)}]{michelstrapp}
\bibinfo{author}{\bibfnamefont{J.~P.~J.} \bibnamefont{{Michels}}}
  \bibnamefont{and} \bibinfo{author}{\bibfnamefont{N.~J.}
  \bibnamefont{{Trappaniers}}}, \bibinfo{journal}{Phys. Lett. A}
  \textbf{\bibinfo{volume}{104}}, \bibinfo{pages}{425} (\bibinfo{year}{1984}).

\bibitem[{\citenamefont{{Hales}}(2005)}]{hales}
\bibinfo{author}{\bibfnamefont{T.~C.} \bibnamefont{{Hales}}},
  \bibinfo{journal}{Ann. Math.} \textbf{\bibinfo{volume}{162}},
  \bibinfo{pages}{1065} (\bibinfo{year}{2005}), \bibinfo{note}{see also T. C.
  Hales, arXiv:math.MG/9811078 (1998).}

\bibitem[{\citenamefont{Musin}()}]{musin}
\bibinfo{author}{\bibfnamefont{O.}~\bibnamefont{Musin}},
  \bibinfo{note}{technical Report, Moscow State University, 2004}.



\bibitem{Sc69}
G. D. Scott and D. M. Kilgour,
{\it Brit. J. Appl. Phys.} {\bf 2} 863 (1969).

\bibitem[{\citenamefont{{Torquato} and {Stillinger}}(2001)}]{category1}
\bibinfo{author}{\bibfnamefont{S.}~\bibnamefont{{Torquato}}} \bibnamefont{and}
  \bibinfo{author}{\bibfnamefont{F.~H.} \bibnamefont{{Stillinger}}},
  \bibinfo{journal}{J. Phys. Chem. B} \textbf{\bibinfo{volume}{105}},
  \bibinfo{pages}{11849} (\bibinfo{year}{2001}).

\bibitem[{\citenamefont{{Torquato} et~al.}(2003)\citenamefont{{Torquato},
  {Donev}, and {Stillinger}}}]{category2}
\bibinfo{author}{\bibfnamefont{S.}~\bibnamefont{{Torquato}}},
  \bibinfo{author}{\bibfnamefont{A.}~\bibnamefont{{Donev}}}, \bibnamefont{and}
  \bibinfo{author}{\bibfnamefont{F.~H.} \bibnamefont{{Stillinger}}},
  \bibinfo{journal}{Int. J. Solids Structures} \textbf{\bibinfo{volume}{40}},
  \bibinfo{pages}{7143} (\bibinfo{year}{2003}).


\bibitem[{\citenamefont{{Donev} et~al.}(2004)\citenamefont{{Donev}, {Torquato},
  {Stillinger}, and {Connelly}}}]{linprogramming}
\bibinfo{author}{\bibfnamefont{A.}~\bibnamefont{{Donev}}},
  \bibinfo{author}{\bibfnamefont{S.}~\bibnamefont{{Torquato}}},
  \bibinfo{author}{\bibfnamefont{F.~H.} \bibnamefont{{Stillinger}}},
  \bibnamefont{and}
  \bibinfo{author}{\bibfnamefont{R.}~\bibnamefont{{Connelly}}},
  \bibinfo{journal}{J. Comp. Phys.} \textbf{\bibinfo{volume}{197}},
  \bibinfo{pages}{139} (\bibinfo{year}{2004}).
  
\bibitem{footnote}
Ref. \cite{torquato15} presents the first exactly solvable disordered sphere-packing
model (``ghost random sequential addition packing) in arbitrary space dimension. Specifically, it
was shown that all of the $n$-particle correlation functions 
of this nonequilibrium model can be obtained analytically for 
all allowable densities and in any dimension. It provides an
exact demonstration of the decorrelation principle; see also
Ref. \cite{decorrelation}

\bibitem[{\citenamefont{{Lubachevsky} and {Stillinger}}(1990)}]{LS1}
\bibinfo{author}{\bibfnamefont{B.~D.} \bibnamefont{{Lubachevsky}}}
  \bibnamefont{and} \bibinfo{author}{\bibfnamefont{F.~H.}
  \bibnamefont{{Stillinger}}}, \bibinfo{journal}{J. Stat. Phys.}
  \textbf{\bibinfo{volume}{60}}, \bibinfo{pages}{561} (\bibinfo{year}{1990}).

\bibitem[{\citenamefont{{Donev}
  et~al.}(2005{\natexlab{c}})\citenamefont{{Donev}, {Torquato}, and
  {Stillinger}}}]{HSalg}
\bibinfo{author}{\bibfnamefont{A.}~\bibnamefont{{Donev}}},
  \bibinfo{author}{\bibfnamefont{S.}~\bibnamefont{{Torquato}}},
  \bibnamefont{and} \bibinfo{author}{\bibfnamefont{F.~H.}
  \bibnamefont{{Stillinger}}}, \bibinfo{journal}{J. Comp. Phys.}
  \textbf{\bibinfo{volume}{202}}, \bibinfo{pages}{737}
  (\bibinfo{year}{2005}{\natexlab{c}}).


\bibitem[{\citenamefont{{Ross} and {Alder}}(1966)}]{melting1}
\bibinfo{author}{\bibfnamefont{M.}~\bibnamefont{{Ross}}} \bibnamefont{and}
  \bibinfo{author}{\bibfnamefont{B.~J.} \bibnamefont{{Alder}}},
  \bibinfo{journal}{Phys. Rev. Lett.} \textbf{\bibinfo{volume}{16}},
  \bibinfo{pages}{1077} (\bibinfo{year}{1966}).

\bibitem[{\citenamefont{{Streett} et~al.}(1974)\citenamefont{{Streett},
  {Raveche}, and {Mountain}}}]{melting2}
\bibinfo{author}{\bibfnamefont{W.~B.} \bibnamefont{{Streett}}},
  \bibinfo{author}{\bibfnamefont{H.~J.} \bibnamefont{{Raveche}}},
  \bibnamefont{and} \bibinfo{author}{\bibfnamefont{R.~D.}
  \bibnamefont{{Mountain}}}, \bibinfo{journal}{J. Chem. Phys.}
  \textbf{\bibinfo{volume}{61}}, \bibinfo{pages}{1960} (\bibinfo{year}{1974}).

\bibitem[{\citenamefont{{Frenkel} and {Smit}}(2002)}]{frenkel}
\bibinfo{author}{\bibfnamefont{D.}~\bibnamefont{{Frenkel}}} \bibnamefont{and}
  \bibinfo{author}{\bibfnamefont{B.}~\bibnamefont{{Smit}}},
  \emph{\bibinfo{title}{{Understanding Molecular Simulation}}}
  (\bibinfo{publisher}{Academic Press}, \bibinfo{year}{2002}).

\bibitem[{\citenamefont{{Salsburg} and {Wood}}(1962)}]{phiJ}
\bibinfo{author}{\bibfnamefont{Z.~W.} \bibnamefont{{Salsburg}}}
  \bibnamefont{and} \bibinfo{author}{\bibfnamefont{W.~W.}
  \bibnamefont{{Wood}}}, \bibinfo{journal}{J. Chem. Phys.}
  \textbf{\bibinfo{volume}{37}}, \bibinfo{pages}{798} (\bibinfo{year}{1962}).

\bibitem[{\citenamefont{Speedy}(1998)}]{speedy}
\bibinfo{author}{\bibfnamefont{R.~J.} \bibnamefont{Speedy}},
  \bibinfo{journal}{J. Phys. Condens. Matter} \textbf{\bibinfo{volume}{10}},
  \bibinfo{pages}{4387} (\bibinfo{year}{1998}).

\bibitem[{\citenamefont{{Chaikin}}(2000)}]{chaikin2}
\bibinfo{author}{\bibfnamefont{P.~M.} \bibnamefont{{Chaikin}}},
  \emph{\bibinfo{title}{{Soft and Fragile Matter, Nonequilibrium Dynamics,
  Metastability and Flow (M. E. Cates and M. R. Evans, editors)}}}
  (\bibinfo{publisher}{Institute of Physics Publishing, London},
  \bibinfo{year}{2000}), chap. \bibinfo{chapter}{Thermodynamics and
  Hydrodynamics of Hard Spheres; the role of gravity}, pp.
  \bibinfo{pages}{315--348}.

\bibitem{footnoteZ}
The specific extrapolation function we use for both $d=4$ and $d=5$ is of the form
\begin{displaymath}
\Delta Z(x) = a_1 x e^{-a_2 x} \cos{(a_3 x + a_4)},
\end{displaymath}
where $a_1, a_2, a_3,$ and $a_4$ are fitting parameters.

\bibitem[{sat()}]{saturated}
\bibinfo{note}{A saturated packing is one that contains no voids large enough
  to accommodate an additional particle; see Ref. 20, for example.}

\bibitem[{no6()}]{no6Disostaticity}
\bibinfo{note}{Due to computational constraints, our packings for $d=6$ were
  produced with a relatively high expansion rate ($\gamma = 10^{-3}$) and were
  not grown to sufficiently high pressures, as necessary to properly
  distinguish between true contacts and near contacts; therefore, we do not
  show results for $d=6$ in this section.}

\end{thebibliography}

\end{document}